\documentclass[runningheads]{llncs}

\usepackage[cmex10]{amsmath}
\usepackage{amssymb}
\usepackage{color}
\usepackage{graphicx}
\usepackage{multirow}
\usepackage{xspace}
\usepackage{subcaption}
\usepackage{cite}
\usepackage{url}
\usepackage{booktabs}
\usepackage{tabularx}
\usepackage{multirow}
\usepackage{enumitem}
\usepackage{textcomp}
\usepackage{bmpsize}
\usepackage{xcolor}
\usepackage{lipsum}

\usepackage{enumitem}
\setlist[enumerate]{itemsep=0mm}

\newcommand{\bp}{\textsc{DeepEvent}\xspace}    
\newcommand{\eg}{\textit{e}.\textit{g}.\@\xspace}
\newcommand{\ie}{\textit{i}.\textit{e}.\@\xspace}
\newcommand{\etal}{\textit{et} \textit{al}.\@\xspace}

\DeclareMathOperator*{\argmax}{arg\,max}

\begin{document}
\title{\large Connecting Web Event Forecasting with Anomaly Detection: A Case Study on Enterprise Web Applications Using Self-Supervised Neural Networks}
\titlerunning{Connecting Web Event Forecasting with Anomaly Detection}
\author{Xiaoyong Yuan\inst{1} \and Lei Ding \inst{2} \and Malek Ben Salem \inst{3} \and Xiaolin Li \inst{4} \and Dapeng Wu \inst{5}}
\authorrunning{Yuan et al.}
\institute{Michigan Technological University, Houghton MI 49931, USA\\ \email{xyyuan@mtu.edu}\and
American University, Washington, DC 20016, USA\\ \email{ding@american.edu}   \and
Accenture Labs, Arlington, VA 22209, USA\\\email{malek.ben.salem@accenture.com} \and
Cognization Lab, Palo Alto, CA 94306, USA \\
\email{xiaolinli@ieee.org}\and
University of Florida, Gainesville, FL 32608, USA\\
\email{dpwu@ufl.edu}
}

\maketitle

\begin{abstract}
Recently web applications have been widely used in enterprises to assist employees in providing effective and efficient business processes. 
Forecasting upcoming web events in enterprise web applications can be beneficial in many ways, such as efficient caching and recommendation.
In this paper, we present a web event forecasting approach, \bp, in enterprise web applications for better anomaly detection.
\bp includes three key features:
web-specific neural networks to take into account the characteristics of sequential web events, self-supervised learning techniques to overcome the scarcity of labeled data, and sequence embedding techniques to integrate contextual events and capture dependencies among web events. 
We evaluate \bp on web events collected from six real-world enterprise web applications. 
Our experimental results demonstrate that \bp is effective in forecasting sequential web events and detecting web based anomalies. 
\bp provides a context-based system for researchers and practitioners to better forecast web events with situational awareness.
\end{abstract}


\section{Introduction}
\label{sec:intro}
Recently web applications play a major role in many enterprises. 
Enterprise web applications boost productivity by assisting employees in performing their daily tasks. 
On one side, web events in the enterprise web applications provide insightful sources for analyzing employee behaviors.
By forecasting web events based on user behavior, we can provide better recommendation, caching, pre-fetching, and load balancing for enterprise web applications~\cite{yang2001mining,el2004fs}.
Web event forecasting has been investigated for many years~\cite{su2000whatnext,awad2012prediction,sharif2018predicting}.
However, two critical challenges exist in characterizing the sequence of web events: 1) Although many works have been conducted in forecasting web events, very few works investigated enterprise web applications. 
Web applications significantly improve productivity in modern enterprises and have become essential components in enterprise operations. 
Enterprise web applications are more well-organized and closely connected to each other compared to other web applications. 
Specific patterns of web events exist in the employees' browsing behavior. 
2) With the increasing complexity and functionality of web applications, it becomes hard to forecast web events. 
Browsing a web page can produce a sequence of web events, and the length of the sequence varies a lot for different web pages. 
For example, a web event is usually made for the HTML of the web page itself, and subsequent events are made for each image, plug-in, audio clip, and other content referenced in the HTML. 
The web events for each kind of content increase the complexity of the sequence. 
The format and semantics of web events vary significantly from application to application. It is urgent to develop techniques to automate the process of characterizing web events and representing them in the desired way.

On the other side, enterprise web applications provide extended connectivity to an organization's assets and increase the attack surface of its web-facing infrastructure. 
Enterprise web applications have become favorite targets of cyber-attacks due to easy access and constantly increasing vulnerabilities. 
Anomaly detection is a critical component to protect web applications against cyber threats. 
Supervised learning based anomaly detection solutions build detection systems by discovering abnormal behavioral patterns with the use of labeled training data~\cite{kruegel2003anomaly,pham2016anomaly,dong2017adaptively,yu2018deephttp}. 
Unfortunately, high quality annotated data is not easy to obtain, given the velocity, volume, and real-time nature of web events. 
Usually, labeled data are very imbalanced as it is hard to collect a large number of labeled anomalies as opposed to normal web events. 
Insufficient and imbalanced training data hinder the performance of machine learning models~\cite{japkowicz2002}. 
Limited labeled data from previous attacks undercut the ability to use supervised models, and constantly evolving attacks make such supervised models irrelevant. 

To address the above challenges, we propose a novel deep neural network based approach, \bp. By connecting web event forecasting and anomaly detection, \bp improves the performance of web event forecasting for complicated web events, while detects anomalies by identifying the most unlikely events in the sequence.
We evaluated \bp on web events collected from six real-world web applications in a company for three months. 

In this paper, we make the following contributions:

\textbf{1) Context-based web events analysis for better user behavior forecasting}:
\bp characterizes not only the individual web events, but also the relationships among the web events that co-occur within a context, \ie, a sequence of events that are commonly produced together for a specific user and web applications' flow characteristics. 
We leverage deep neural networks to model the context of a web event in a sequence.
Based on the context, \bp predicts the web events that are likely to appear with associated probabilities. 

\textbf{2) Self-supervised learning to overcome the need for labeled data}:
By connecting web event forecasting and anomaly detection, we can formulate anomaly detection as a self-supervised learning problem~\cite{raina2007self,jing2019self}.
Self-supervised learning tasks do not require any prior knowledge of anomalous web events' features and their sequential relationships, but leverage naturally existing evidence as labeled data for training. Therefore, self-supervised learning overcomes the need for labeled data in anomaly detection. In addition, we leverage a pre-training process that learns the representations of web events to further compensate for the lack of labeled data.

\textbf{3) Quantitative measures of the anomaly}:
Different from traditional anomaly detection methods, \bp goes beyond binary prediction. Upon encoding the relationship between web events within the vector space, we can quantitatively predict the web events appearing in similar contexts. Therefore, \bp predicts if a web event belongs to normal behavior and also measures the deviation of the abnormal event with respect to expected normal events. In such a way, we provide a quantitative measure of the detected anomaly. 

We empirically validate that forecasts made by \bp are more accurate than baseline solutions. We show that with the assist of web event forecasting, \bp can identify anomalous events (\eg, real-world exploits, widely-used web attacks) using the proposed quantitative measures. We find these results encouraging and note that they highlight the benefit of forecasting web events that co-occur within a context for anomaly detection. 

\section{Workflow of \bp}
\label{sec:workflow}
\begin{figure}[!tb]
\centering
\includegraphics[width=\linewidth]{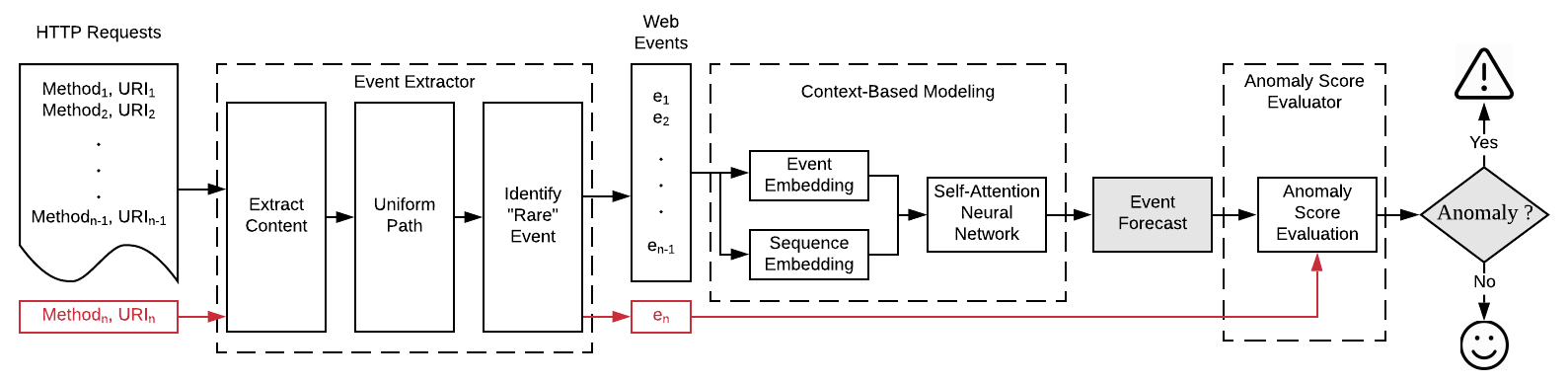}
\caption{Workflow of \bp.}
\label{fig:workflow}
\vspace{-20pt}
\end{figure}

We design three main components in \bp: an event extractor to extract semantic events from web requests, a context-based model for sequential web event forecast as well as an anomaly score evaluator. 





Figure~\ref{fig:workflow} illustrates the workflow of \bp. 
\bp ingests a sequence of web requests from web applications as its input, \eg, with a length of $n-1$. 
Then it extracts the contents from web requests and converts them into a sequence of events $\{e_1, e_2, \ldots, e_{n-1}\}$. 
After event extraction, \bp performs context-based modeling where it first encodes the events into event embedding and sequence embedding. Event embedding represents the content of each event. Sequence embedding represents the order of each event in the sequence. 
The embedding output will then be used to build a neural network (as described in Section~\ref{subsec:contex-based-model}) to learn long-term dependencies of the events. The trained context-based model will calculate the probability distribution of possible events to appear as the next one given $\{e_1, e_2, \ldots, e_{n-1}\}$ and provide the forecast of the upcoming event $e_n$. In addition, \bp uses the predicted probability to evaluate the anomaly score of $e_n$.  

We describe the three main components in the following sections.

\subsection{Event Extraction}
\label{sec:data_pre}
The purpose of event extraction is to extract semantic events from web requests. We use sequences of events to characterize web requests. For well-formatted web requests (\eg, following REST API design), one can easily map these requests to events by extracting their HTTP methods and well-defined endpoints. However, most web requests may not share a well-organized representation or follow a consistent format. For example, URI paths may not be named around resources. Web developers may use various URI paths for the same resource on the web server. For instance, WordPress provides five different URIs and a custom one for users to access a post\footnote{\url{https://codex.wordpress.org/Using_Permalinks}}. URI paths can be generated by randomized algorithms or encoding algorithms. Diverse web requests impede the effort to extract semantic events from URIs.

We propose a three-step event extraction method, including content extracting, path uniforming, and ``rare'' event identifying. 

\begin{itemize}[leftmargin=*]
\item \textbf{Extract Content}. We extract three components from web requests: HTTP methods (\ie, GET, POST, UPDATE, etc.), URI paths, and the number of URI parameters. Our experimental results show that using merely these three components are effective in representing user behavior.

\item \textbf{Uniform Path}. We apply a two-character Markov Chain model to detect the ``random'' elements in URI paths. We first segment URI paths into ``elements'' separated by special characters such as ``/'' and ``-''. We then investigate every character in the element from left to right. If the likelihood of the upcoming character based on the preceding two characters is lower than a certain threshold, then we consider the element as ``random.''~\footnote{We detect randomness in URIs based on a gibberish detection tool (\url{https://github.com/rrenaud/Gibberish-Detector}).}

\item \textbf{Identify ``RARE'' Events}. We consider the events occurring less than $T$ times in the training data as ``RARE'' events. In this way, we learn the information of ``RARE'' events during training, which helps us to understand if such rare events are anomalous or not.
    
\end{itemize}

The entire process of pre-processing data proceeds as follows:
\begin{enumerate}[leftmargin=*]
    \item Extract HTTP methods, URI paths, and URI queries from web requests;
    \item Segment URI paths into ``elements'' by special characters as delimiters;
    \item Flag the ``elements'' as ``RANDOM,'' if they are randomly generated or encoded;
    \item Calculate the number of key-value pairs in the URI query;
    \item Concatenate HTTP method, ``derandomized'' URI path, and the number of the URI query as an event.
    \item If an event has never seen in the training set or has occurred less than $T$ times, convert the event into a ``RARE'' event.
\end{enumerate}


\subsection{Context-Based Modeling}
\label{subsec:contex-based-model}
We propose context-based web request modeling, which takes a sequence of contextual events as input and outputs a sequence of corresponding events. We can mask the event of interest in the input sequence and train a model to predict it for the given sequence. 
Recurrent Neural Networks (RNNs), as well as their variants such as Long Short-Term Memory (LSTM)~\cite{lstm}, 
have been proposed for security analytics in the sequential analysis due to their outstanding performance. Recently, self-attention neural networks ~\cite{devlin2018bert} have been shown to be much more effective to capture long-term dependencies in a sequence compared to RNNs. 
Specifically, RNN passes the hidden states through the previous state while self-attention neural networks construct direct links between events within the context, which brings great merit in learning from the long-distance context. 

\textbf{Self-supervised learning.}
\label{sec:pretrain}
To compensate for the lack of labeled data, we design a self-supervised learning task for event forecasting. 
Most existing supervised models are limited to the high-quality labeled data. 
In this paper, we leverage the existing event requests as the labels without any manual  annotations. 
In practice, we randomly mask 25\% events in the input sequence and replace these masked events with ``mask'' labels in the input sequence. 
We train the neural networks to predict ``mask'' events. 
In this way, the neural network learns the relationship of events and their dependency in the sequences.
We use this neural network as a pretrained model for further event forecasting and anomaly detection.
Our experimental results show that self-supervised pre-training significantly improves the prediction performance.

\subsection{Anomaly Detection}
\label{sec:detection}
We propose a way to calculate the anomaly score to quantitatively measure the likelihood of a new web request being anomalous. Given the current context, we predict a set of web requests that are likely to appear with associated probabilities. For a received web request, we rank it with the predicted set of web requests based on their associated probabilities calculated by the trained neural network. We calculate the anomaly score for the incoming web request as follows:
\begin{equation}
\label{eq:anomaly_score}
 s = 1-\frac{1}{\tau + 1},
\end{equation}
where $\tau$ denotes the rank of the newly received request based on its likelihood to appear (\ie, the probability calculated by the context-based model), and $s$ is in the range of $(0, 1)$. Higher anomaly scores indicate higher confidence level in classifying the event as an anomaly. 

The anomaly score indicates the degree of deviation of the incoming web request from the expected normal requests. In Section~\ref{sec:anomaly_eval}, we show that the proposed anomaly score is able to differentiate anomalous web requests (\eg, produced by various real-world web based attacks) from normal requests.


\section{Methodology of Context-Based Modeling}
\label{sec:methods}
In this section, we describe the detailed approach for context-based modeling (Section~\ref{subsec:contex-based-model}) and explain how we adapt three different types of neural networks to predict web requests. We compare the performance of the three neural networks in Section~\ref{sec:network_compare}.

\subsection{Self-Attention Based Modeling}

\begin{figure}[!tb]
\centering
\includegraphics[width=0.6\linewidth]{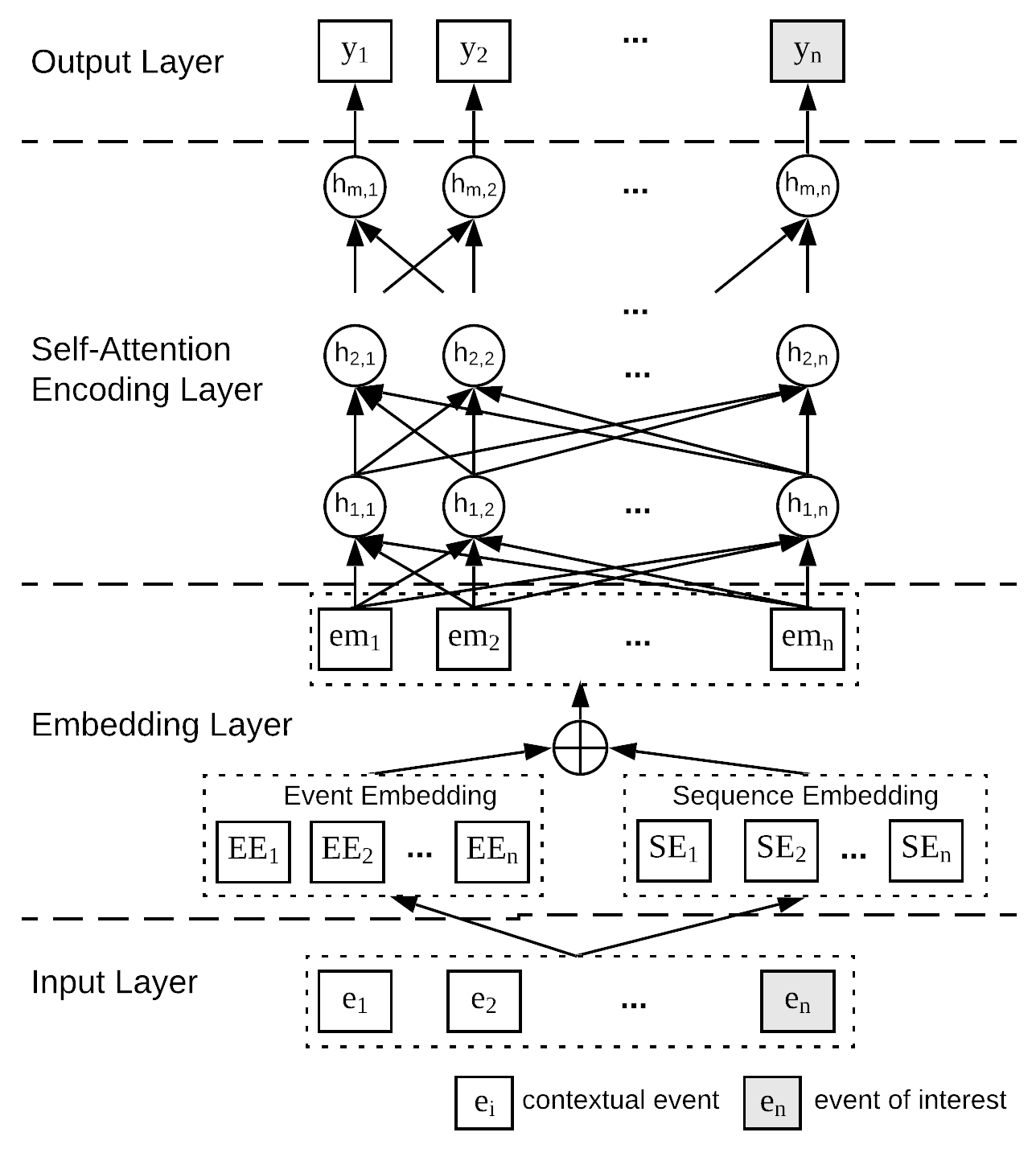}
\caption{Self-Attention Based Modeling.}
\label{fig:single_model}
\vspace{-20pt}
\end{figure}

The design of self-attention based modeling is shown in Figure~\ref{fig:single_model}. We first embed input events using the embedding layer, then use a self-attention neural network to encode the sequence and learn the dependency between events. In the output layer, we apply a Softmax function to squash the neural network and predict future events with associated probabilities. 

We introduce two critical components of our adaptation: 1) event embedding and sequence embedding; and 2) a self-attention neural network.

\textbf{Event embedding and sequence embedding.}
Word embedding is widely used to represent the semantic meaning of tokens in NLP tasks. We adopt the basic idea of word embedding by extracting semantic representation of web events and generate event embedding. Sequential information represents the relative positions of events in the sequence. However, self-attention neural networks do not contain sequential information of events, because the distances of events are the same. Therefore, we add sequential information into the neural networks using a sequence embedding layer. Specifically, our embedding layer maps the web event and its position in the sequence into two 128-dimension vectors: $\{EE_1, EE_2, \ldots, EE_n\}$ and $\{SE_1, SE_2,$ $\ldots, SE_n\}$. After generating the embedding on events and their positions, we sum up these two sequences as an embedded sequence $\{em_1, em_2, \ldots, em_n\}$, and feed it to the encoding layers of the self-attention network. 

\textbf{Self-attention neural network.}
We use a new type of neural networks, self-attention neural network~\cite{vaswani2017attention}, to solve sequential prediction problem. Specifically, we adapt BERT~\cite{devlin2018bert}, a self-attention based neural network. BERT has been shown to outperform RNNs in almost all NLP tasks, and achieve state-of-the-art performance~\cite{devlin2018bert}. 
The success of BERT mainly comes from a scaled dot-product attention neural network:
\begin{equation}
\mathrm{Attention}(Q, K, V) = \mathrm{Softmax}(\frac{QK^T}{\sqrt{d_k}})V,
\end{equation}
where $Q$, $K$, $V$ are the query, key, and value of dimension $d_k$. In the self-attention neural network of \bp, $Q$, $K$, $V$ come from the same sequence of embedded events. The network learns to pay attention to the specific events in the sequence and captures the dependencies between events in the sequence. We use a model called Transformer, including a multi-head attention neural network (stacking several self-attention neural networks), a normalization layer, and a Softmax function. For the details of self-attention mechanism and Transformer, we refer the reader to~\cite{vaswani2017attention}.
\vspace{-0.3cm}
\subsection{Bi-LSTM Based Modeling}
\begin{figure}[!tb]
\centering
\begin{subfigure}{0.47\linewidth}
\centering
\includegraphics[width=\linewidth]{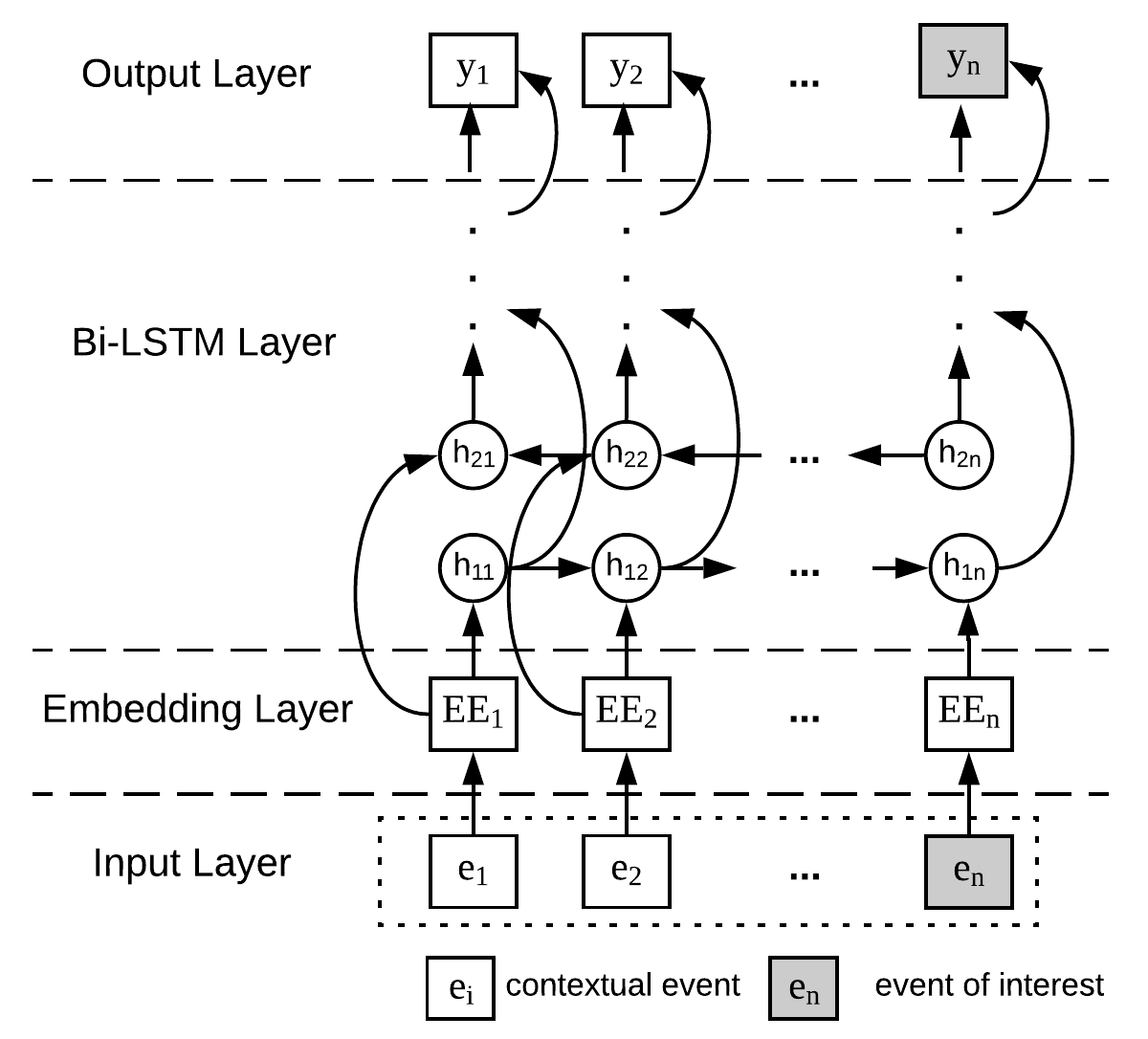}
\caption{Bi-LSTM Based Modeling.}
\label{fig:lstm_model}
\end{subfigure}
\begin{subfigure}{0.47\linewidth}
\centering
\includegraphics[width=\linewidth]{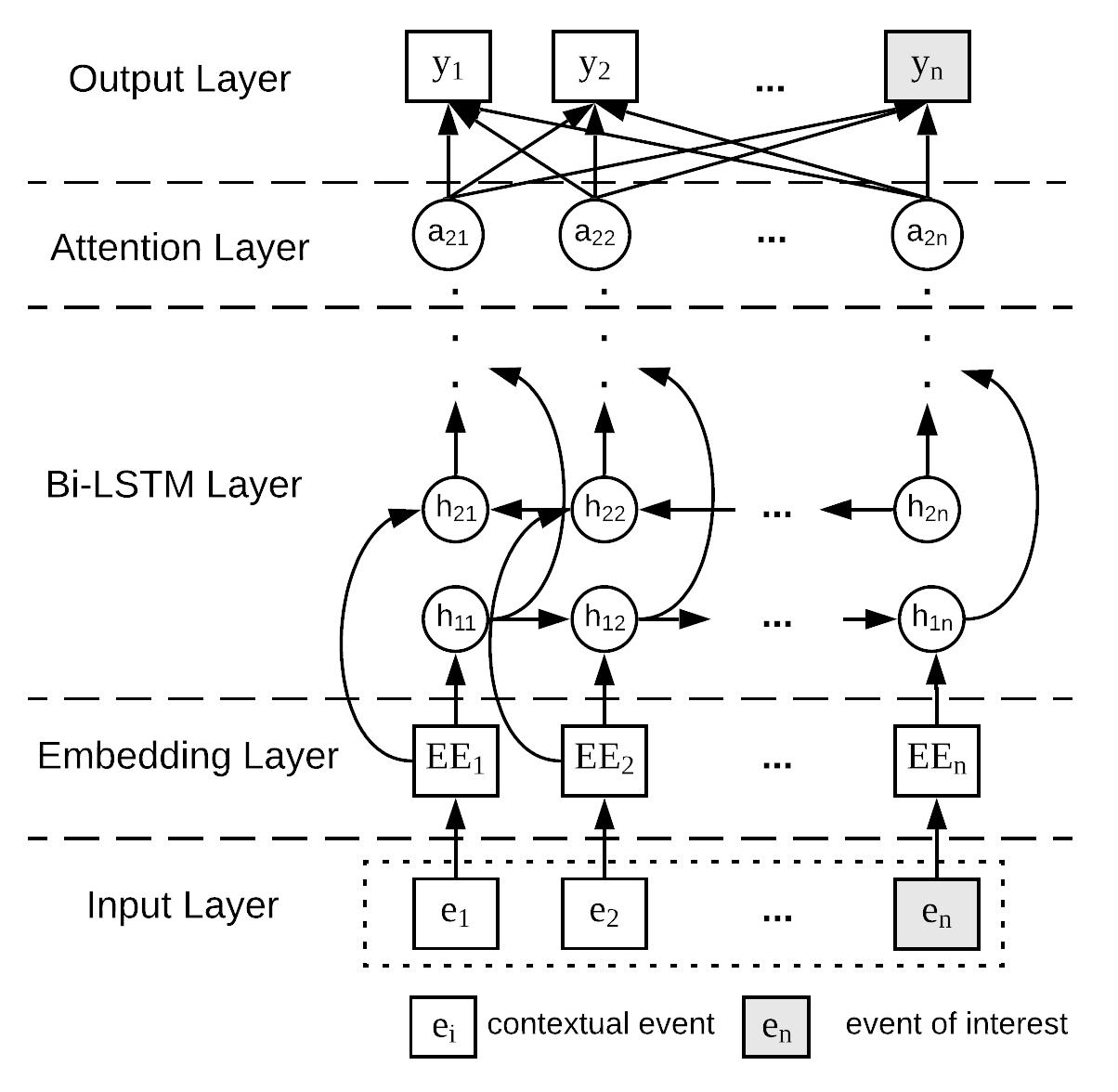}
\caption{LSTM-Attention Based Modeling.}
\label{fig:attention_model}
\end{subfigure}
\caption{Bi-LSTM and LSTM-Attention Based Modeling.}
\vspace{-20pt}
\end{figure}

\vspace{-0.3cm}
In addition to self-attention neural network, we implement a bidirectional LSTM neural network, called Bi-LSTM, which is widely used in sequential analysis. 
Similarly to Self-Attention Neural Network, we apply the event embedding here to encode events and use LSTM to represent their sequential relationship. 
Figure~\ref{fig:lstm_model} illustrates the architecture of Bi-LSTM based modeling. 
Bi-LSTM uses an embedding layer to convert an input event into a 128-dimensional vector, then deploys multiple bidirectional LSTM layers to extract semantic information from the events. 
We test 1, 2, 3 layers of LSTM for each direction in the experiment. A fully connected layer with a Softmax function is stacked on top of LSTM layers to output the final prediction related to the event of interest. 

\subsection{LSTM-Attention Based Modeling}
Attention mechanism is recently proposed to surpass recurrent neural networks by remembering longer sequences. 
The attention mechanism aims to pinpoint key events from a long sequence. We adapt an LSTM-Attention neural network using an additive attention mechanism. Figure~\ref{fig:attention_model} illustrates the architecture of LSTM-Attention based modeling. The additive attention mechanism sums up the outputs of LSTM with their weights and outputs the weighted sum as the prediction. The LSTM-Attention neural network consists of an embedding layer, bidirectional LSTM layers, an additive attention layer, and a fully connected layer with Softmax. The embedding layer and LSTM layers follow the setting in Bi-LSTM neural network. The attention layer learns the weight of each event and applies them to the final output. 
Note that the neural network in~\cite{yu2018deephttp} detects anomaly based on the content of a single request. However, we use LSTM-Attention to predict an event in a sequence.
We set the hidden number of the embedding layer in Bi-LSTM and LSTM-Attention to 128 and apply a drop-out mechanism with 20\% dropout in LSTM layers to avoid overfitting. 
\vspace{-0.3cm}

\section{Evaluation}
\label{sec:eval}
\vspace{-0.3cm}
In this section, we describe the experiments to evaluate \bp on real-world web applications. We introduce our experimental settings in Section~\ref{sec:setting}. We designed experiments to answer the following questions:
\begin{enumerate}[leftmargin=*]
    \item What is \bp's performance in web event forecast compared with existing methods (Section~\ref{sec:model_compare})?
    \item How effective is \bp in evaluating real-world threats (Section~\ref{sec:anomaly_eval})?
    \item How do different neural networks perform in \bp (Section~\ref{sec:network_compare})?
    \item How do different model settings (\eg, pre-training, window size) affect prediction performance (Section~\ref{sec:setting_compare})?
\end{enumerate}


\subsection{Experimental Setup}
\label{sec:setting}

\begin{table}[!tb]
\centering
\caption{Dataset Description. 
}
\label{tab:dataset}
\begin{tabular}{lrrrr}
\toprule
\multirow{2}{*}{\textbf{Application}} & \multicolumn{3}{c}{\textbf{\# of Events}} & \multirow{2}{*}{\shortstack{\textbf{\# of Unique} \\ \textbf{Events}}} \\ \cline{2-4}
                             & \textbf{Train}      & \textbf{Valid}    & \textbf{Test}     &                                \\ \hline
Workqueue                    & 146,489   & 18,026      & 19,004  & 101                                  \\ 
DataRepo1                         & 98,798    & 11,500      & 13,596  & 258                                  \\ 
DevOpsApp                      & 134,230   & 23,652      & 35,517  & 1,787                                \\ 
DataAnalyzer1                           & 757,626   & 106,836     & 64,681  & 1,442                                \\ 
DataAnalyzer2                       & 363,787   & 59,285      & 65,222  & 329                                  \\ 
DataRepo2                           & 63,862    & 7,235       & 8,307   & 37                                   \\ \bottomrule
\end{tabular}
\end{table}

\begin{figure}[!tb]
\centering
\includegraphics[width=0.95\linewidth]{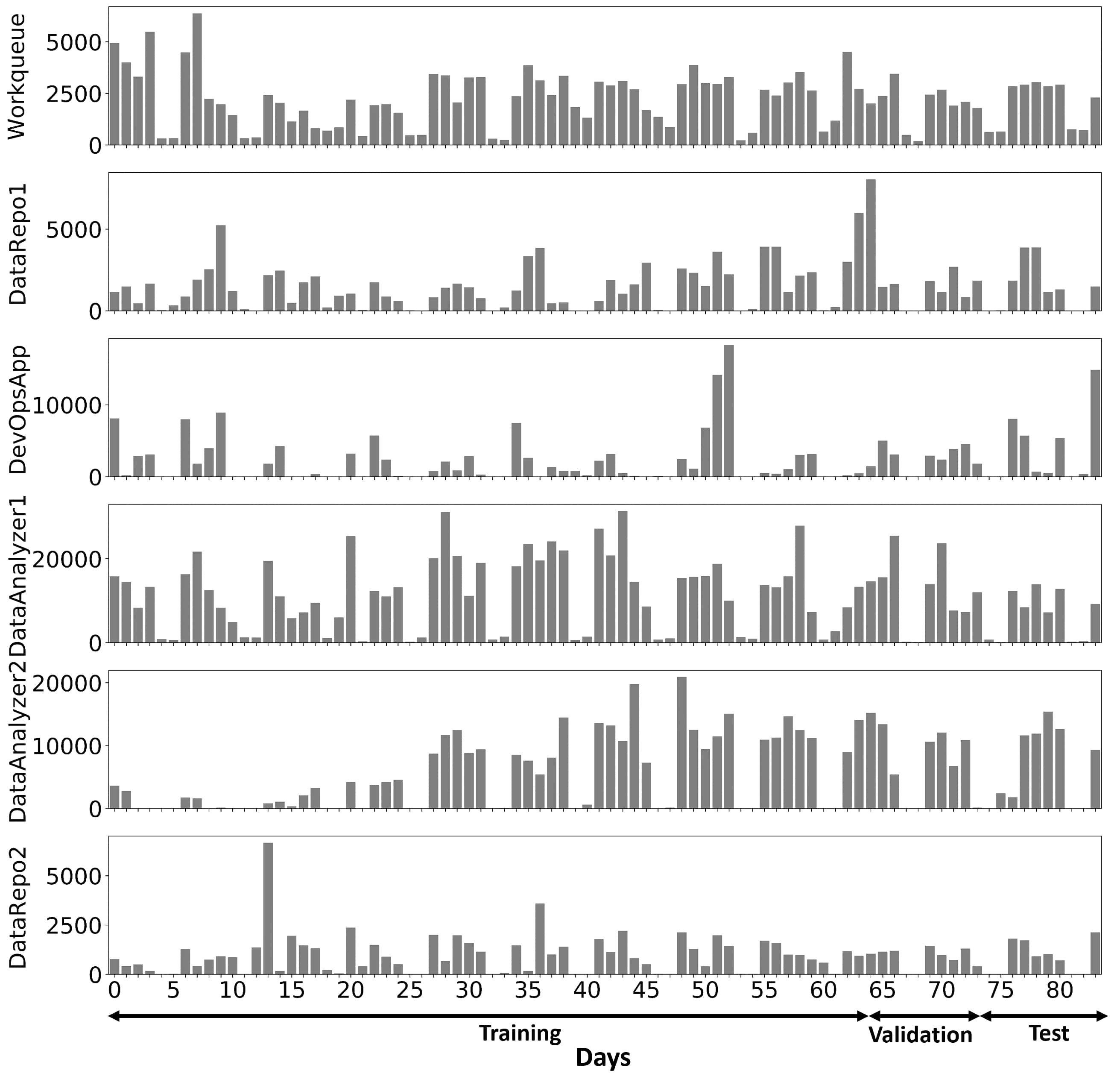}
\caption{Number of HTTP Requests of Six Web Applications in the Dataset. We annotate the days used in the training/validation/test data. 
The number of HTTP requests on weekdays is higher than that on weekends, indicating the activity pattern of an enterprise network. }
\label{fig:data_day}
\vspace{-15pt}
\end{figure}

\textbf{Dataset.} 
We evaluate \bp on six real-world enterprise web applications described: 1) Workqueue: work queue of data collections and analysis;2) DataRepo1: meta data repository; 3) DevOpsApp: continuous integration application for DevOps; 4) DataAnalyzer1: data reporting and visualization; 5) DataAnalyzer2: data analysis and visualization; 5) DataRepo2: graph and meta data repository. We collected 84 days of HTTP requests from the above applications in a real-world enterprise network. 
Figure~\ref{fig:data_day} shows the number of HTTP requests for each web application in the dataset. 
We separate the dataset based on the date and use the first 64 days of data for training, 10 days for validation, and the last 10 days for testing. We assume all the HTTP requests are legitimate. Note that we only use the HTTP requests produced by normal users, and we exclude HTTP requests that are automatically generated by machines (\eg, heartbeat requests). 
Table~\ref{tab:dataset} summarizes the statistics of training/validation/test data of those six applications. We report the number of web events used in training/validation/test data and the number of unique events observed in the training data. 
The number of unique events indicates how many different events after event extraction occur in the datasets.

\textbf{Evaluation metrics.}
In the experiments, we evaluate the performance of \bp using Top-N Accuracy and Anomaly Score. \textbf{Top-N Accuracy} measures the event prediction performance. \textit{Top-N Accuracy} calculates the percentage of the correct event occurs among the top N events predicted by the model. In the experiments, we report Top-1 and Top-10 accuracy. \textbf{Anomaly Score}, as defined in Equation~\ref{eq:anomaly_score}, evaluates the capability in terms of differentiating anomalous requests from normal requests. 


\textbf{Modeling settings.}
We use self-attention based modeling in \bp in the experiments in Section~\ref{sec:model_compare} and Section~\ref{sec:anomaly_eval} as it performs better in general compared to Bi-LSTM, and LSTM-attention based modeling for the six enterprise web applications. We show and discuss the comparison among the three for all six web applications in  Section~\ref{sec:network_compare}. 

We set the threshold of ``rare events'' to 2 ($T=2$). We set the number of hidden layers of \bp to $8$, the number of attention heads in each layer to $8$, the number of hidden neurons in each head to $128$, the batch size to $128$. We use Cross-Entropy as our loss function. We optimize the loss function using Adam~\cite{kingma2014adam} with L2 weight decay. We set the learning rate to $0.001$ in pre-training and reduce the learning rate by $10$ in training.
We train the models for 100 epochs. 
To accelerate the training process, we adopt an early-stop strategy, which ceases training if the cross-entropy loss of the validation data does not decrease in the past 10 epochs. 


\vspace{-0.5cm}
\subsection{Evaluation of \bp on Web Event Forecast}
\label{sec:model_compare}
\vspace{-0.2cm}
In this section, we evaluate \bp on the web event forecast. 
We compare \bp with two baseline models: a Markov model and an N-gram model.

The \textbf{Markov model} predicts the upcoming event depending on the current event. It assumes the upcoming event does not depend on the previous event. It learns transition probability from the training data. In this paper, we use a first-order Markov model~\cite{kemeny1976markov}.

The \textbf{N-gram model} aims to provide maximum likelihood estimates for the last event ${e_N}$ given previous $N-1$ contextual events $\{e_1, e_2, \ldots, e_{N-1}\}$~\cite{yang2001mining}:
\begin{equation}
    \argmax Pr\{e_N|e_1, e_2, \ldots, e_{N-1}\},
\end{equation}
where $N$ denotes the number of events considered for prediction. Following the same setting in~\cite{shenccs18}, we use a $3$-gram model to predict the upcoming events.

\begin{table}[!tb]
\centering
\caption{Model Comparison. We compare the performance of Markov model, $3$-gram model, and our proposed \bp model.}
\label{tab:baseline_compare}
\begin{tabular}{llrr}
\toprule
\textbf{Application}           & \textbf{Method}    & \textbf{Top-1 Accuracy (\%)} & \textbf{Top-10 Accuracy (\%)}  \\ \hline
\multirow{3}{*}{Workqueue}     & Markov Model       & 56.30          & 91.24  \\
                               & 3 Gram Model       & 61.90          & 94.17  \\ 
                               & \textbf{\bp} & \textbf{75.21}          & \textbf{99.27}         \\ \hline
\multirow{3}{*}{DataRepo1}     & Markov Model       & 62.59          & 95.73  \\  
                               & 3 Gram Model       & 74.17          & 96.76   \\ 
                               & \textbf{\bp} & \textbf{79.53}          & \textbf{97.82}   \\ \hline
\multirow{3}{*}{DevOpsApp}     & Markov Model       & 44.84          & 80.42  \\ 
                               & 3 Gram Model       & 48.01          & 82.34  \\  
                               & \textbf{\bp} & \textbf{56.73}          & \textbf{89.51}       \\ \hline
\multirow{3}{*}{DataAnalyzer1} & Markov Model       & 55.10          & 86.53  \\  
                               & 3 Gram Model       & 63.07          & 90.90 \\  
                               & \textbf{\bp} & \textbf{70.10}          & \textbf{95.44}     \\ \hline
\multirow{3}{*}{DataAnalyzer2} & Markov Model       & 70.21          & 95.96\\ 
                               & 3 Gram Model       & 73.21          & 97.10\\ 
                               & \textbf{\bp} & \textbf{78.45}          & \textbf{97.72}\\ \hline
\multirow{3}{*}{DataRepo2}     & Markov Model       & 95.05          & 99.74\\  
                               & 3 Gram Model       & 95.56          & 99.79\\  
                               & \textbf{\bp} & \textbf{97.47}          & \textbf{99.87}\\ \bottomrule
\end{tabular}
\vspace{-20pt}
\end{table}

In our experiments, we observe that \bp performs better than the Markov and the $3$-gram models. We report Top-1 and Top-10 accuracy of the three methods in Table~\ref{tab:baseline_compare}. 
We highlight all the best results among the three methods. 
For all six applications, \bp increases 11.2\% Top-1 accuracy and 3.48\% Top-10 accuracy on average compared with the best results of the Markov model and $3$-gram Model. 

The $3$-gram model performs better than the Markov model, and \bp outperforms both of them. 
This is because the long-distance context is lost in Markov model and $3$-gram model. In contrast, \bp has longer-term memory compared with the other two models so it can capture the sequential relationships of events that are not adjacent to each other, which is especially important for web applications because task-critical web requests may be separated by ``not so relevant'' requests. For example, browsing a web page can produce a sequence of web requests, where one web request is made for the HTML of the web page itself, and subsequent requests are made for each image, plug-in, audio clip, and other content referenced in the HTML. The web requests for each piece of content increase the length of the sequence. These make task-critical requests located far from each other. 
In general Top-10 accuracy is better than Top-1 accuracy because it is more likely to provide a correct prediction with more candidate events. 
\vspace{-0.5cm}
\subsection{Evaluation of \bp on Anomaly Detection}
\label{sec:anomaly_eval}
\vspace{-0.1cm}
In this section, we evaluate \bp's capability in distinguishing normal web requests from different types of anomalous requests such as real-world exploits, popular web attacks, and randomly injected HTTP requests. We compare their anomaly scores evaluated by \bp with normal requests.

\textbf{Real-world exploits}: We investigate five real-world exploits on DevOpsApp, which is built on top of Jenkins (\url{jenkins.io/}): CVE-2016-9299, CVE-2016-0792, CVE-2018-1999001, CVE-2018-1999002, CVE-2019-1003000.
We mix the requests generated by the exploits with normal ones and test how \bp evaluates the exploits.

\textbf{Web attacks}: 
We investigate five widely-used attacks against web applications: SQL Injection~\cite{sql_injection}, Cross-site Scripting (XSS)~\cite{xss}, Buffer Overflow~\cite{buffer_overflow}, CRLF Injection~\cite{crlf_injection}, Server-Side Includes (SSI) Injection~\cite{ssi_injection}. We mix the requests produced by those attacks with normal ones in the same way as real-world exploits and test how \bp evaluates such requests. 
We use OWASP Zed Attack Proxy (ZAP)~\cite{zap}, to actively scan and attack web applications. ZAP is one of the most popular open-source tools for web security and vulnerability assessment. 
ZAP accesses the web application using a normal user's credentials. It first crawls all the URIs of the web application and then crafts malicious web requests to exploit the vulnerabilities. We collect all the malicious requests and categorize them by the type of attacks. 
Then we randomly inject them into normal web requests to evaluate \bp.

\textbf{Random injection}: We conduct experiments to test how \bp evaluate requests produced by abnormal behaviors, such as requests are generated by normal users but at abnormal occurrences. We simulate arbitrary web requests based on normal requests and inject them into a sequence of normal requests. 

\begin{table}[!tb]
\centering
\begin{minipage}[t]{0.45\textwidth}
\caption{Anomaly Score Evaluation for DevOpsApp.}
\label{tab:jenkins-attack}
\begin{tabular}{lr}
\toprule
\textbf{Request Type}  & \begin{tabular}[r]{@{}r@{}}\textbf{Average} \\\textbf{Anomaly Score}\end{tabular} \\ \hline
Normal Requests      & 0.316                 \\ 
CVE-2019-1003000     & 0.996                 \\ 
CVE-2016-9299        & 0.996                 \\ 
CVE-2016-0792        & 0.996                 \\ 
CVE-2018-1999001     & 0.787                 \\ 
CVE-2018-1999002     & 0.996                 \\ 
SQL Injection        & 0.976                 \\ 
Cross-site Scripting & 0.975                 \\ 
Buffer Overflow      & 0.964                 \\ 
CRLF                 & 0.964                 \\ 
SSI Injection        & 0.964                 \\ 
Random Injection     & 0.995                 \\ \bottomrule
\end{tabular}
\end{minipage}
\qquad
\begin{minipage}[t]{0.45\textwidth}
\caption{Average Anomaly Score Comparison between Normal Requests and Randomly Injected Requests.}
\label{tab:inject}
\begin{tabular}{lrr}
\toprule
\textbf{Application}  & \begin{tabular}[r]{@{}r@{}}\textbf{Normal}\\ \textbf{Request}\end{tabular} & \begin{tabular}[r]{@{}r@{}}\textbf{Random} \\ \textbf{Injection}\end{tabular} \\ \hline
Workqueue     & 0.151           & 0.948                    \\ 
DataRepo1     & 0.132           & 0.978                    \\ 
DevOpsApp     & 0.316           & 0.995                    \\ 
DataAnalyzer1 & 0.197           & 0.995                    \\ 
DataAnalyzer2 & 0.140           & 0.980                    \\ 
DataRepo2     & 0.014           & 0.887                    \\ \bottomrule
\end{tabular}
\end{minipage}
\vspace{-20pt}
\end{table}
Table~\ref{tab:jenkins-attack} shows the performance of anomaly detection for DevOpsApp. All real exploits achieve high anomaly scores, compared to normal requests (0.316). CVE-2018-1999001 gains the lowest anomaly score (0.787). Yet, it still has a large distance from normal requests. Four real exploits (CVE-2016-9299, CVE-2016-0792, CVE-2018-1999002, CVE-2019-1003000) were identified as ``RARE'' events. ``RARE'' events gain a high anomaly score (0.996), which suggests EventExtractor performs well in extracting web events from HTTP requests. 







In our experiments, anomaly scores calculated by \bp perform well at distinguishing the requests generated by web attacks from normal requests. 
In general, our proposed solution can distinguish anomalous requests from the normal request. For instance, the malicious requests and normal requests can be distinguished with a threshold value of 0.9 - most malicious requests have an anomaly score higher than 0.9 while most normal requests have a score lower than 0.9. We observe a few false positives for normal requests due to the low frequency of these normal events, which might be further improved with more training data to recognize unexpected normal requests.
Some web attacks may gain a slightly high false negative rate (being classified as normal requests) than others. For example, Buffer Overflow has 2.37\% false positive rate, because Buffer Overflow usually targets URI parameters as well as payload, which is not covered in this paper. 



Table~\ref{tab:inject} shows the average anomaly score of normal requests and randomly injected requests for six web applications. Normal requests get low anomaly scores (0.158 on average) while randomly injected ones could raise the alarm with an extremely high anomaly score (0.964 on average). The large performance gap indicates that we can set a threshold to differentiate normal requests from randomly injected requests.



\vspace{-0.5cm}
\subsection{Neural Network Comparison}
\label{sec:network_compare}
\vspace{-0.2cm}
We compare three different neural networks proposed in context-based modeling in \bp: Bi-LSTM, LSTM-attention, and Self-attention. To make a fair and comprehensive comparison, we use the same settings for three neural networks. We evaluate Bi-LSTM and LSTM-attention networks with three different numbers of LSTM layers: 1, 2, 3, and evaluate all three neural networks with five different window sizes: 8, 16, 32, 64, 128 and report the best results.

\begin{table}[!tb]
\centering
\caption{Comparison Between Three Neural Networks.}
\label{tab:model_comparison}
\begin{tabular}{llrr}
\toprule
\textbf{Application}           & \textbf{Neural Network} & \textbf{Top-1 Accuracy (\%)} & \textbf{Top-10 Accuracy (\%)}\\ \hline
\multirow{3}{*}{Workqueue}     & Bi-LSTM                 & 74.69          & 99.00 \\ 
                               & LSTM-Attention          & 73.55          & 98.66\\ 
                               & Self-Attention          & \textbf{75.21}          & \textbf{99.27}\\ \hline
\multirow{3}{*}{DataRepo1}     & Bi-LSTM                 & 78.20          & 97.36 \\ 
                               & LSTM-Attention          & 78.61          & 96.39\\ 
                               & Self-Attention          & \textbf{79.53}          & \textbf{97.82} \\\hline
\multirow{3}{*}{DevOpsApp}     & Bi-LSTM                 & 53.06          & 87.52  \\ 
                               & LSTM-Attention          & 51.58          & 84.63  \\ 
                               & Self-Attention          & \textbf{56.73}          & \textbf{89.51}  \\\hline
\multirow{3}{*}{DataAnalyzer1} & Bi-LSTM                 & 70.84          & 93.47 \\ 
                               & LSTM-Attention          & \textbf{71.51}          & 93.15\\
                               & Self-Attention          & 70.10          & \textbf{95.44}\\ \hline
\multirow{3}{*}{DataAnalyzer2} & Bi-LSTM                 & 78.23          & 97.21\\
                               & LSTM-Attention          & 77.75          & 97.05 \\ 
                               & Self-Attention          & \textbf{78.45}          & \textbf{97.72} \\\hline
\multirow{3}{*}{DataRepo2}     & Bi-LSTM                 & 97.21          & \textbf{99.91} \\ 
                               & LSTM-Attention          & 97.45          & 99.90\\ 
                               & Self-Attention          & \textbf{97.47}          & 99.87\\ \bottomrule
\end{tabular}
\vspace{-10pt}
\end{table}

\begin{figure*}[!tb]
\centering
\begin{subfigure}{0.32\linewidth}
\centering
\includegraphics[width=\linewidth]{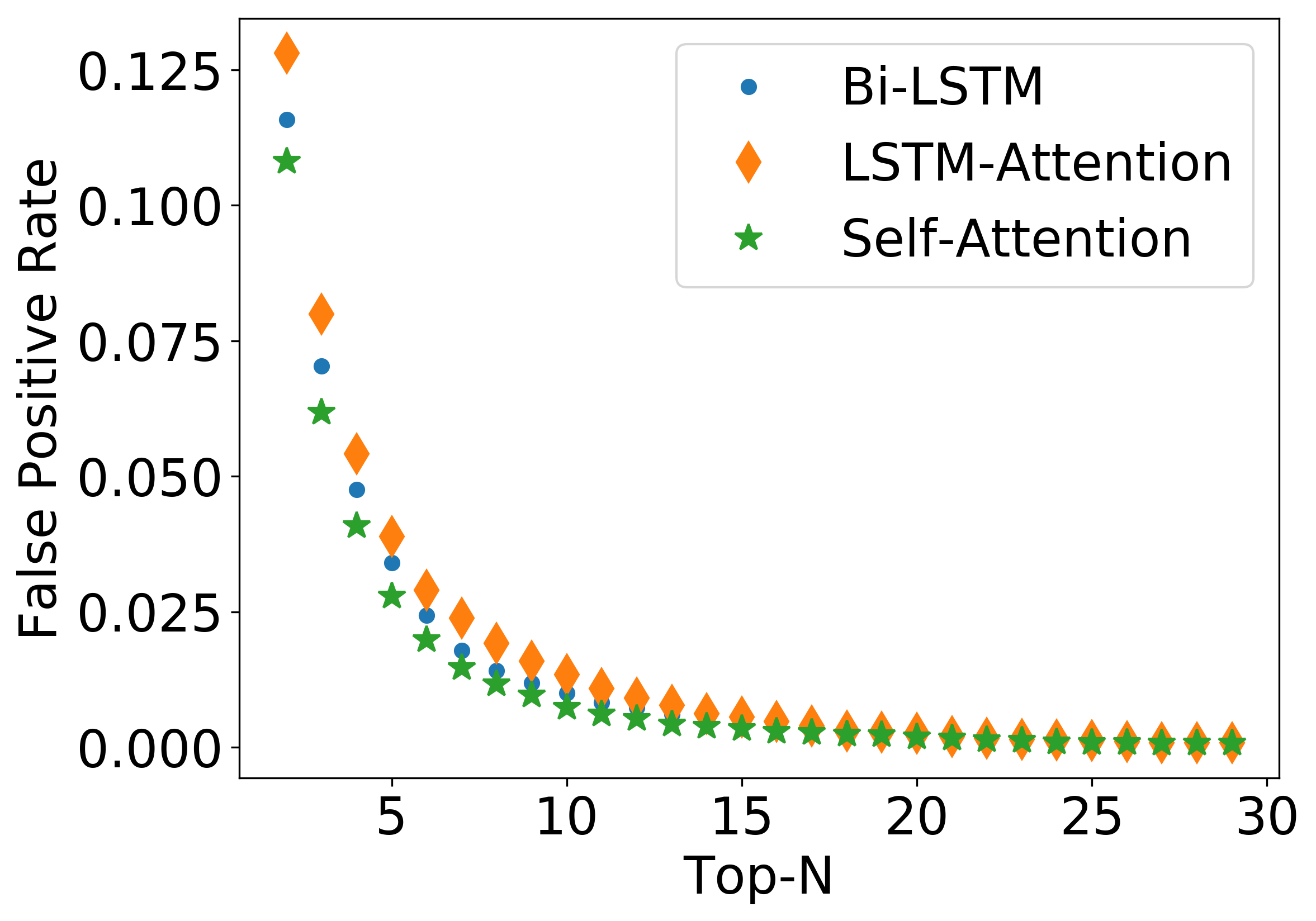}
\caption{Workqueue}
\end{subfigure}
\begin{subfigure}{0.32\linewidth}
\centering
\includegraphics[width=\linewidth]{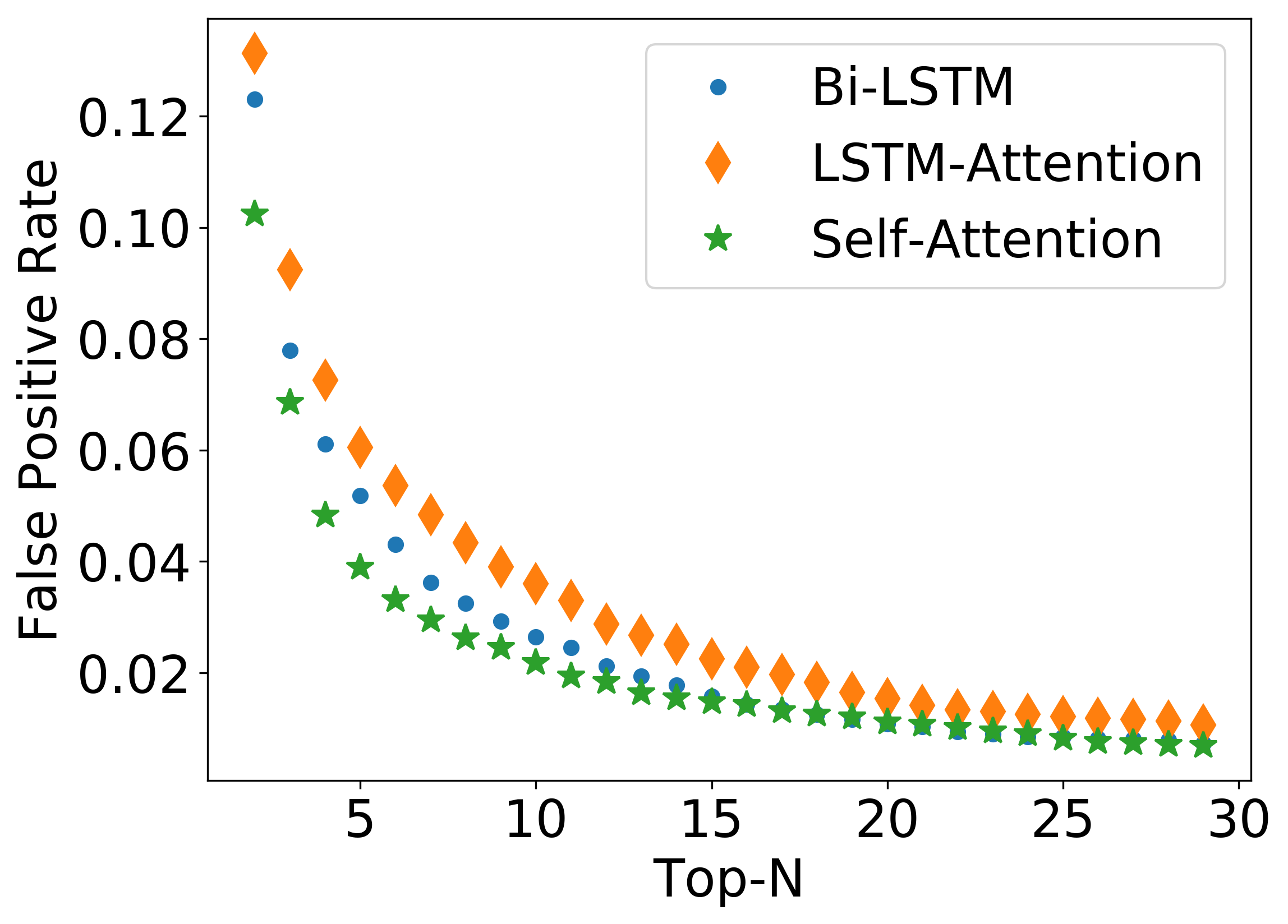}
\caption{DataRepo1}
\end{subfigure}
\begin{subfigure}{0.32\linewidth}
\centering
\includegraphics[width=\linewidth]{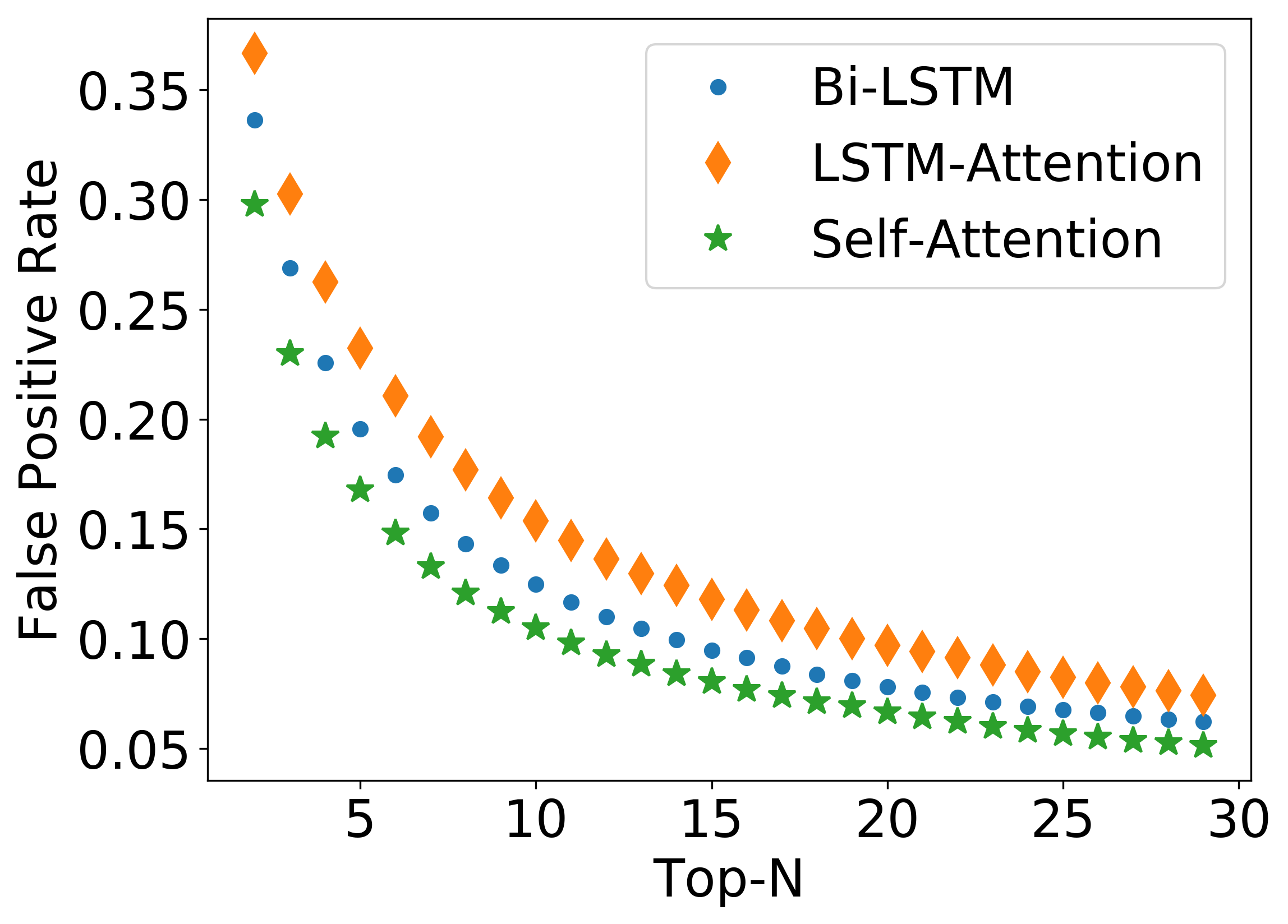}
\caption{DevOpsApp}
\end{subfigure}
\begin{subfigure}{0.32\linewidth}
\centering
\includegraphics[width=\linewidth]{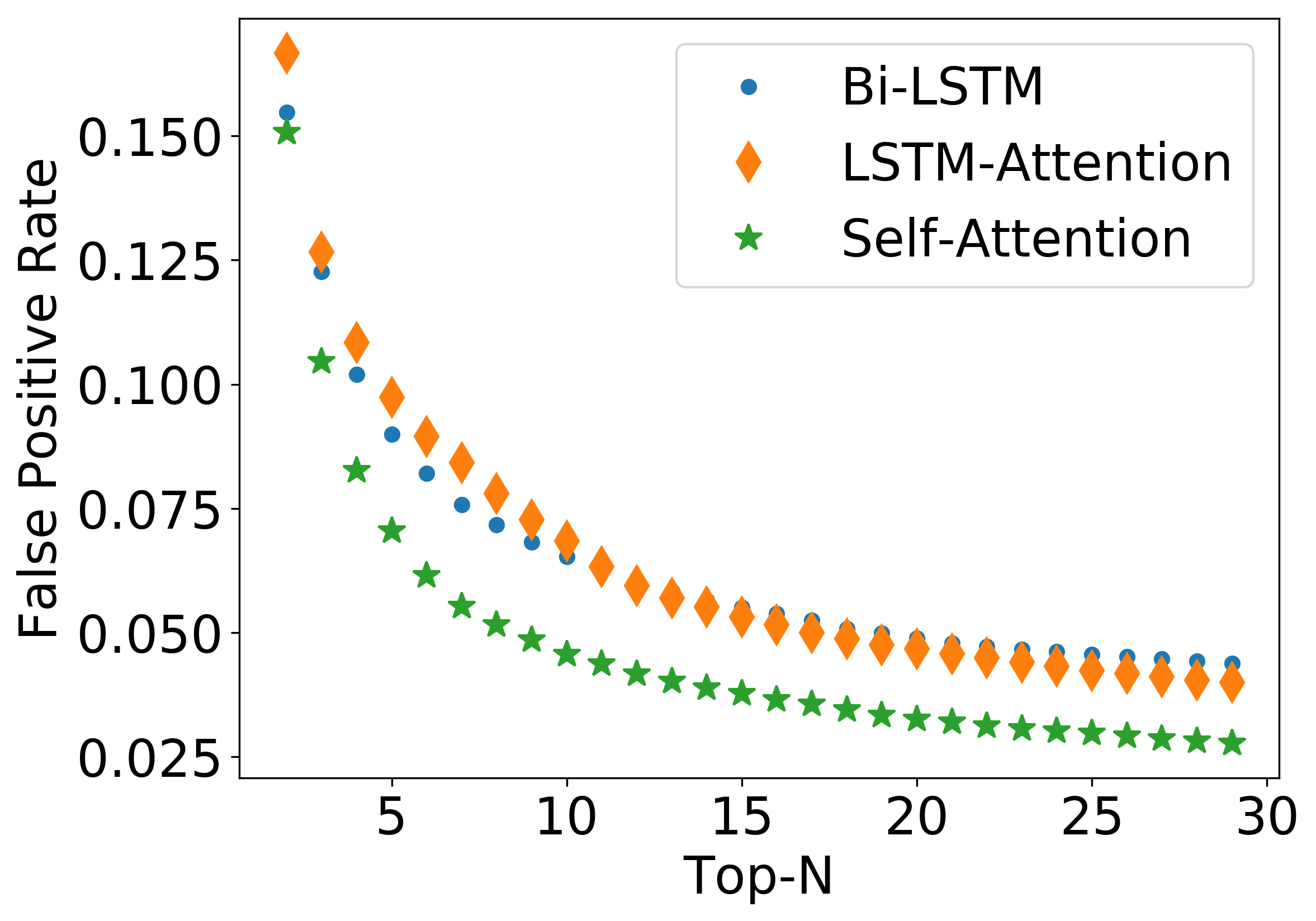}
\caption{DataAnalyzer1}
\end{subfigure}
\begin{subfigure}{0.32\linewidth}
\centering
\includegraphics[width=\linewidth]{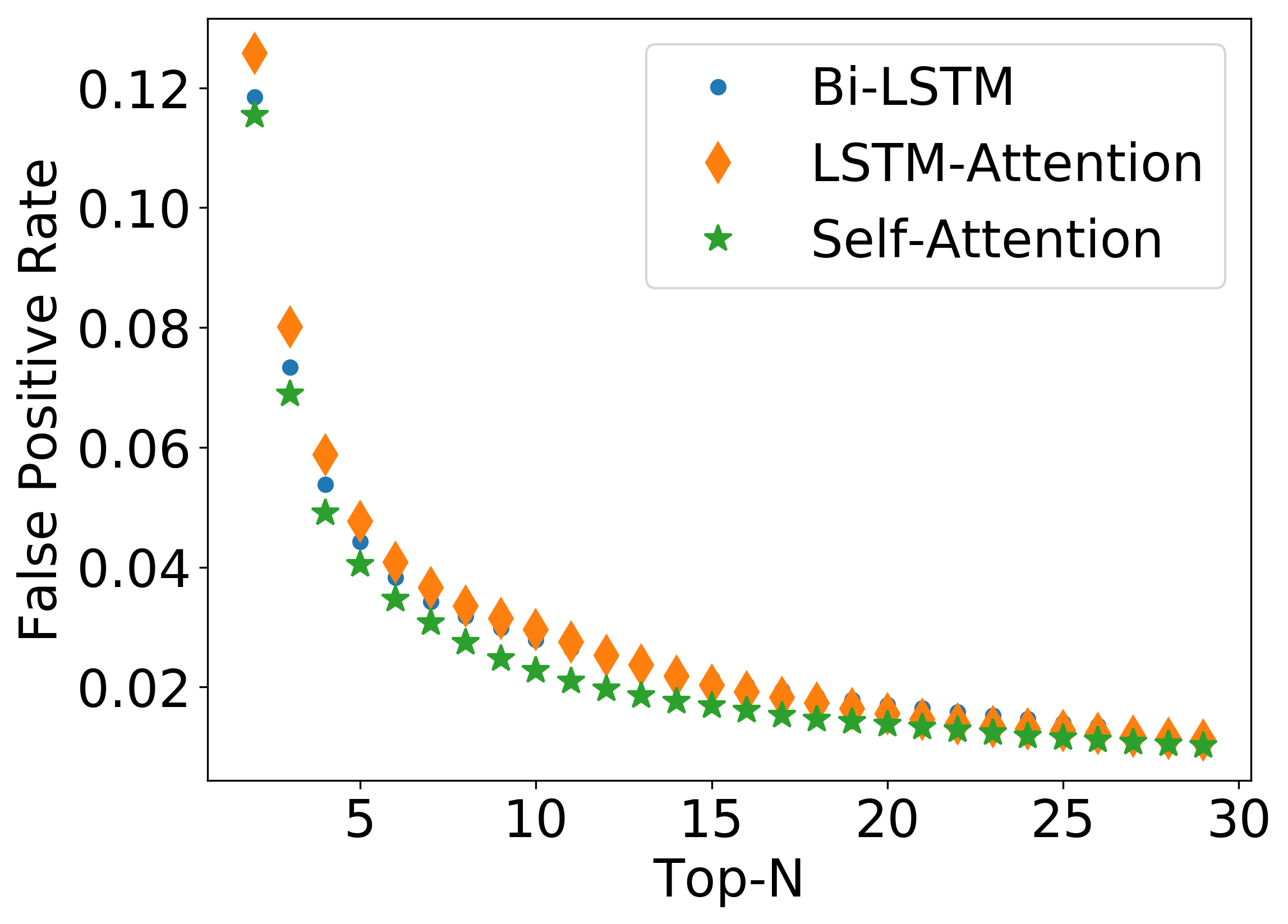}
\caption{DataAnalyzer2}
\end{subfigure}
\begin{subfigure}{0.32\linewidth}
\centering
\includegraphics[width=\linewidth]{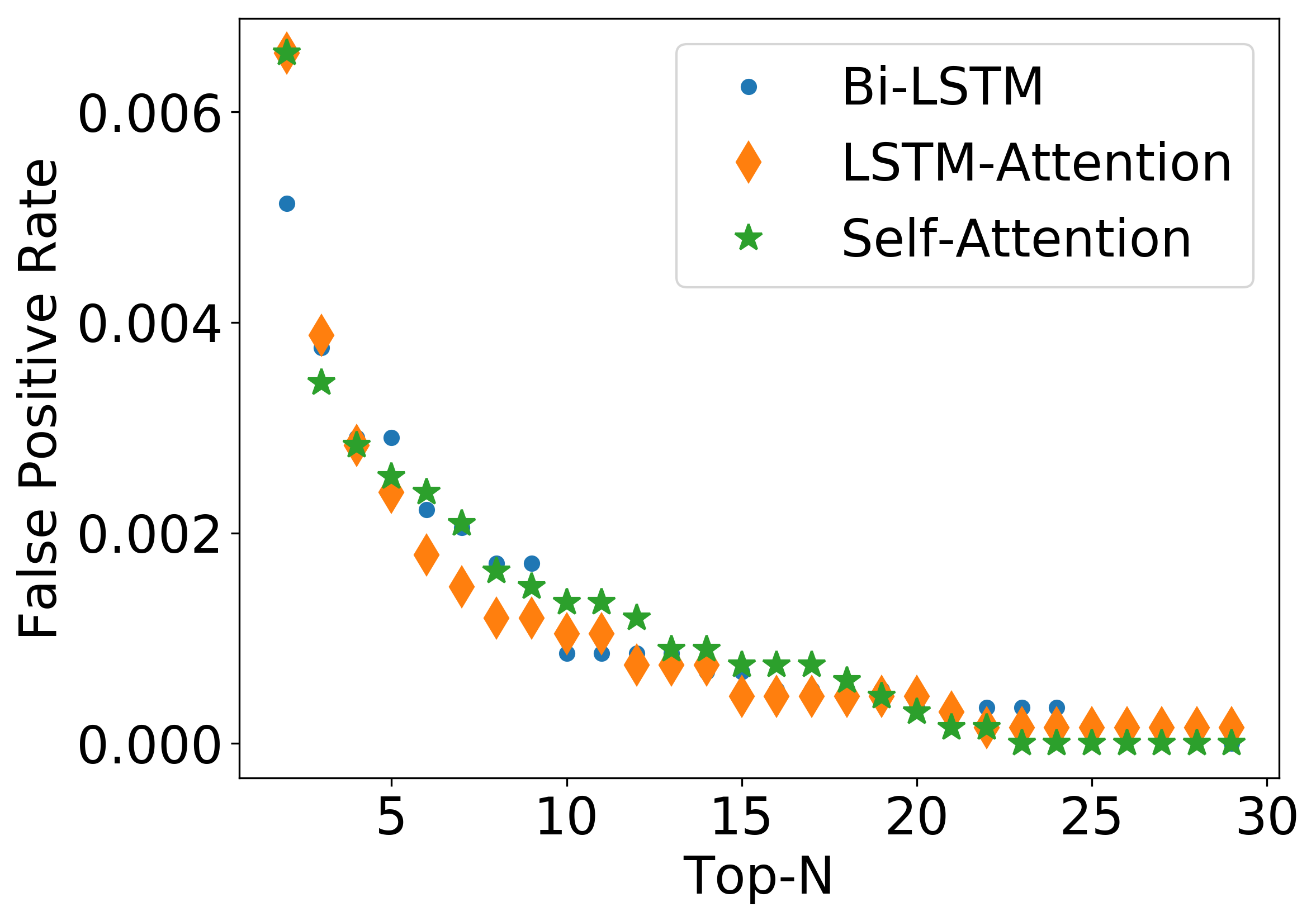}
\caption{DataRepo2}
\end{subfigure}
\caption{False Positive Rate (FPR) Comparison: We report FPR of Bi-LSTM, and LSTM-Attention with different alarm thresholds (Top-N). 
}
\label{fig:fpr_online}
\vspace{-20pt}
\end{figure*}

In Table~\ref{tab:model_comparison}, we report the best results of three neural networks with different settings (window size, number of LSTM layers, with/without pre-training). From the experimental results, we observe that Self-Attention usually achieves the highest accuracy (Top-1, Top-10). In most cases, Bi-LSTM performs second best. 
Bi-LSTM and LSTM-attention networks need to pass the hidden states through a long path to learn from the long-distance context. The context might be lost in a long path because of gradient vanishing~\cite{hochreiter2001gradient}. This may not be suitable for web applications that have task-critical requests located far from each other. 
On the other hand, the self-attention neural network constructs direct links between requests within the context, which brings great merit in learning from the long-distance context. 


To evaluate false positive and false negative rates of the three neural networks, we flag an event as an abnormal event if it is not among the top $K$ candidate events predicted by the neural network. 
Correct predictions are considered as true positives. 
We select the best setting of the three neural networks (Bi-LSTM, LSTM-Attention, and Self-Attention) and calculate their false positive rates based on different alarm thresholds using Top-N, \ie, threshold 10 means if the event is not in Top-10 prediction, it will be labeled as an alarm.
As illustrated in Figure~\ref{fig:fpr_online}, false positive rate decreases when we use a large threshold $K$. For the same threshold, Self-Attention based model achieves lower false positive rates than the other two models for all web applications except for DataRepo2. For DataRepo2, all three neural networks achieve extremely low false positive rates, less than 1\%. 
\vspace{-0.4cm}

\subsection{Evaluation of Different Model Settings}
\label{sec:setting_compare}
In this section, we evaluate the impact of different model settings on prediction performance.


\begin{figure*}[!tb]
\centering
\begin{subfigure}{0.32\linewidth}
\centering
\includegraphics[width=\linewidth]{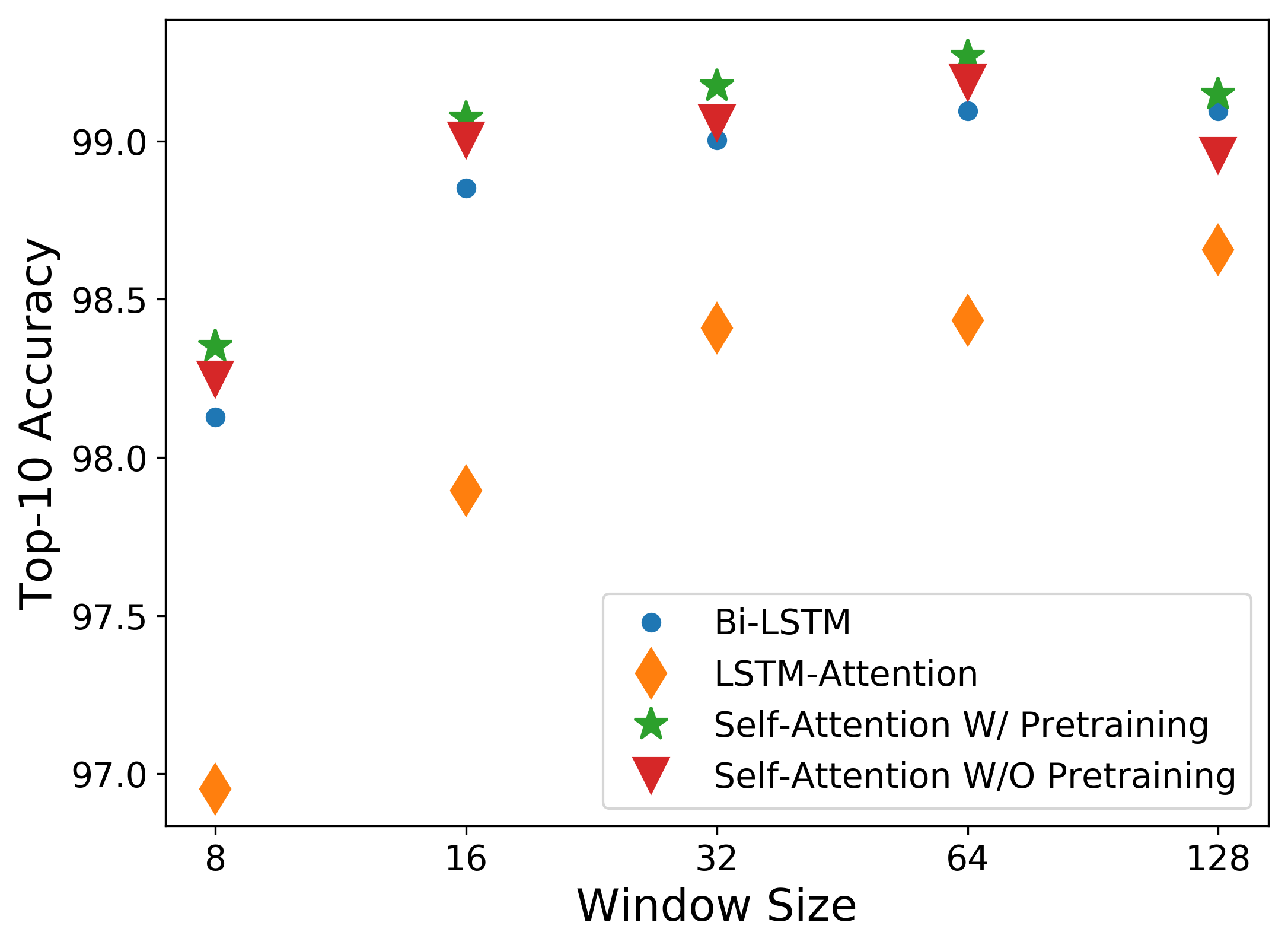}
\caption{Workqueue}
\end{subfigure}
\begin{subfigure}{0.32\linewidth}
\centering
\includegraphics[width=\linewidth]{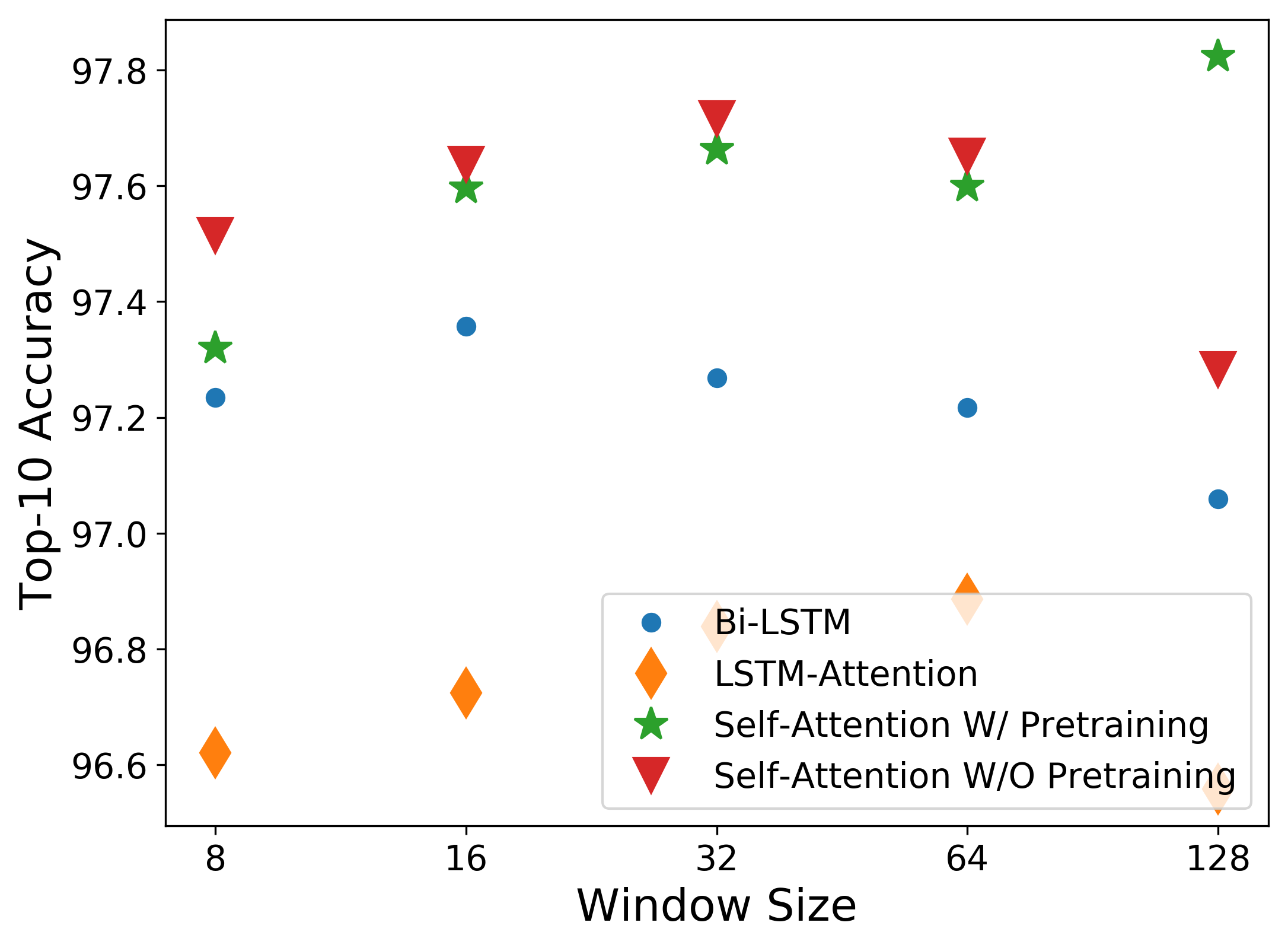}
\caption{DataRepo1}
\end{subfigure}
\begin{subfigure}{0.32\linewidth}
\centering
\includegraphics[width=\linewidth]{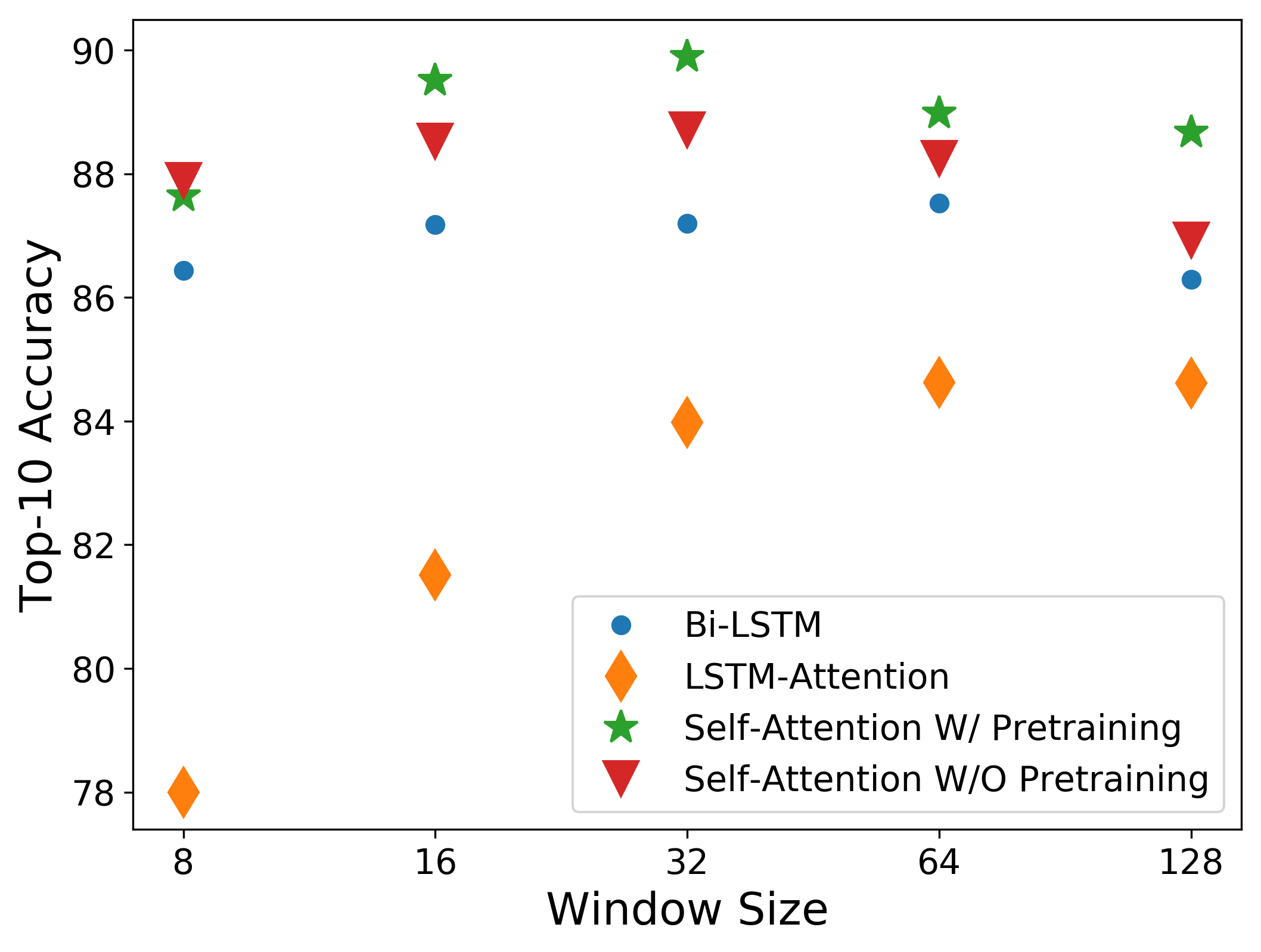}
\caption{DevOpsApp}
\end{subfigure}
\begin{subfigure}{0.32\linewidth}
\centering
\includegraphics[width=\linewidth]{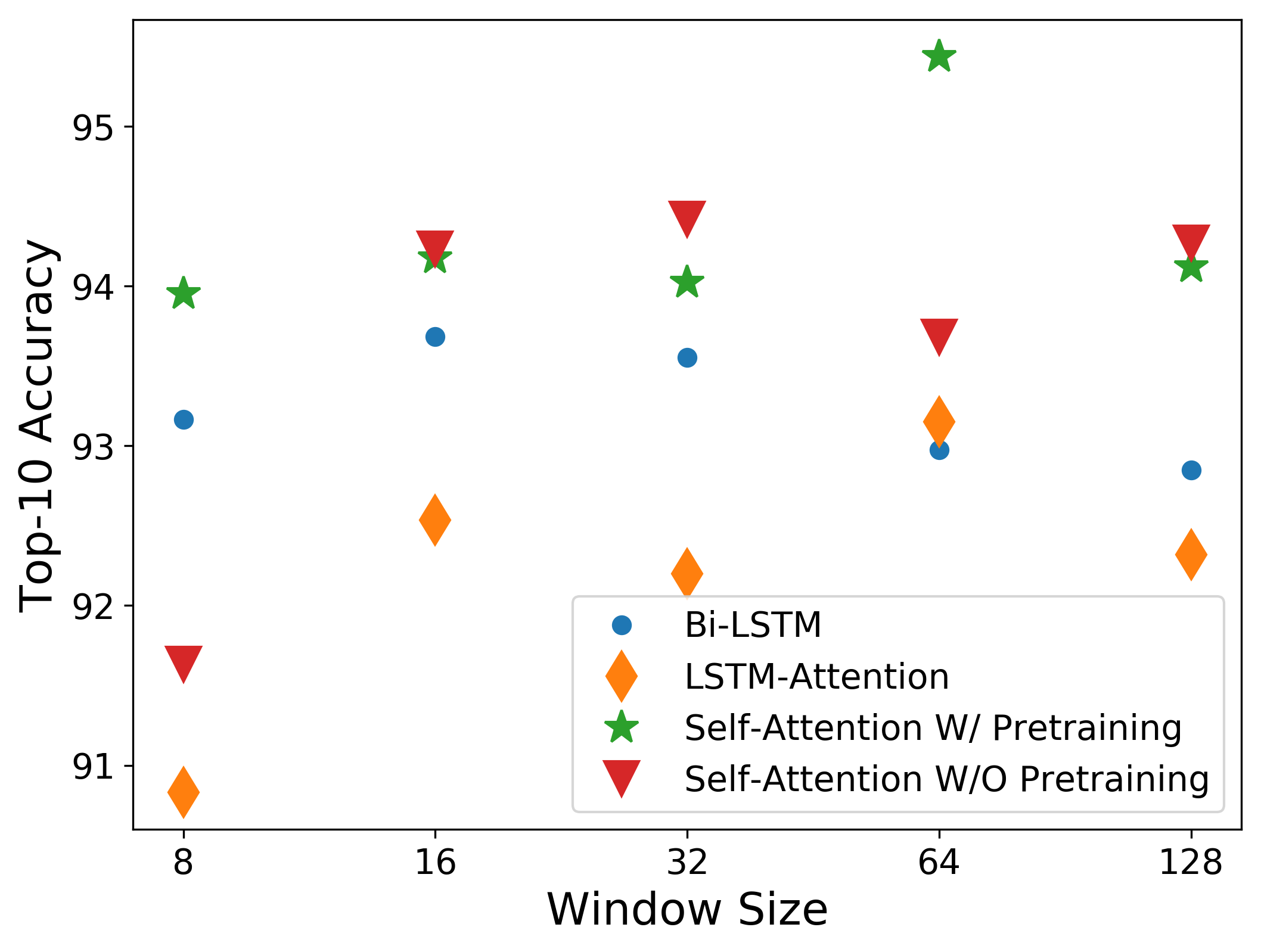}
\caption{DataAnalyzer1}
\end{subfigure}
\begin{subfigure}{0.32\linewidth}
\centering
\includegraphics[width=\linewidth]{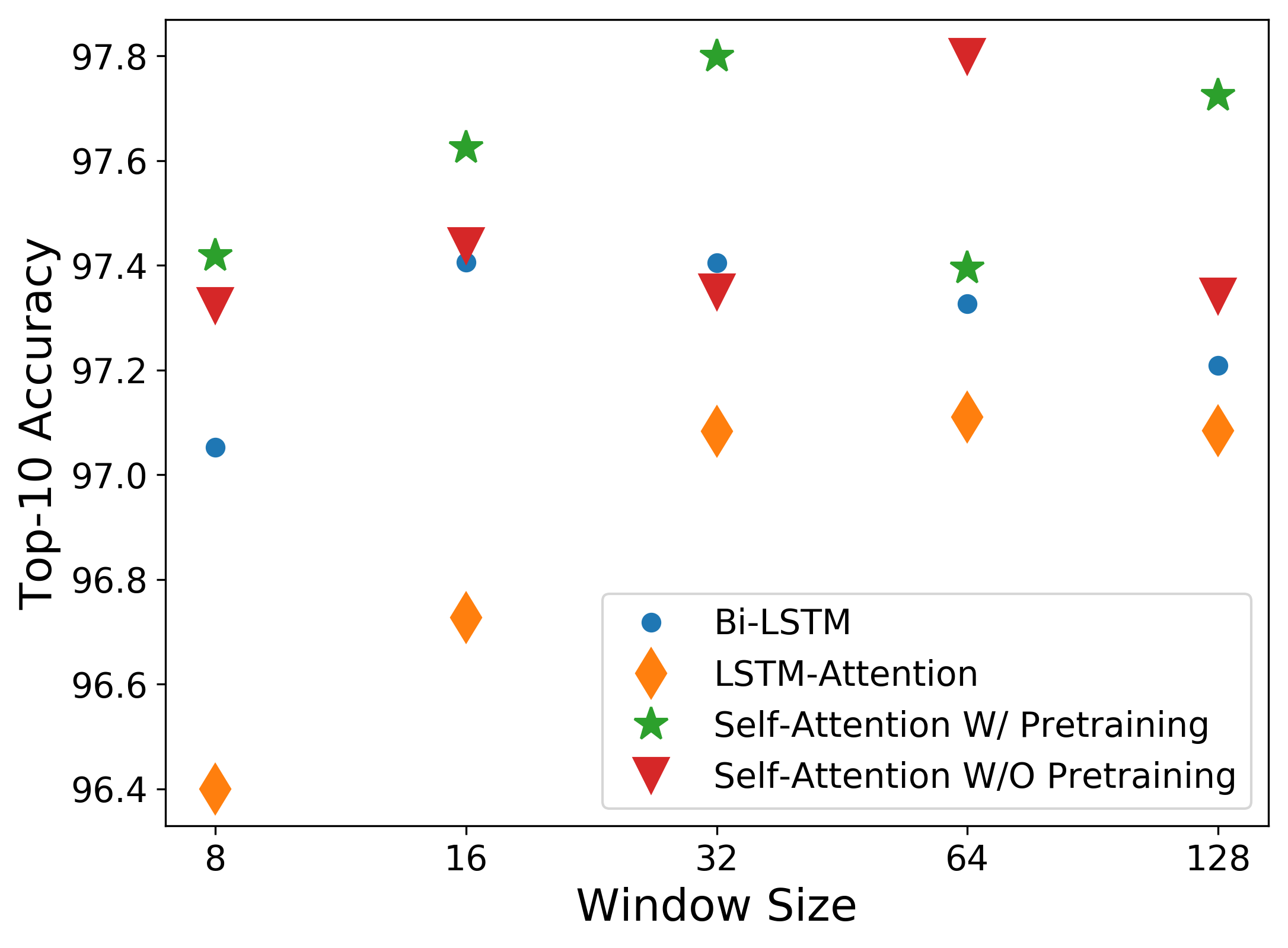}
\caption{DataAnalyzer2}
\end{subfigure}
\begin{subfigure}{0.32\linewidth}
\centering
\includegraphics[width=\linewidth]{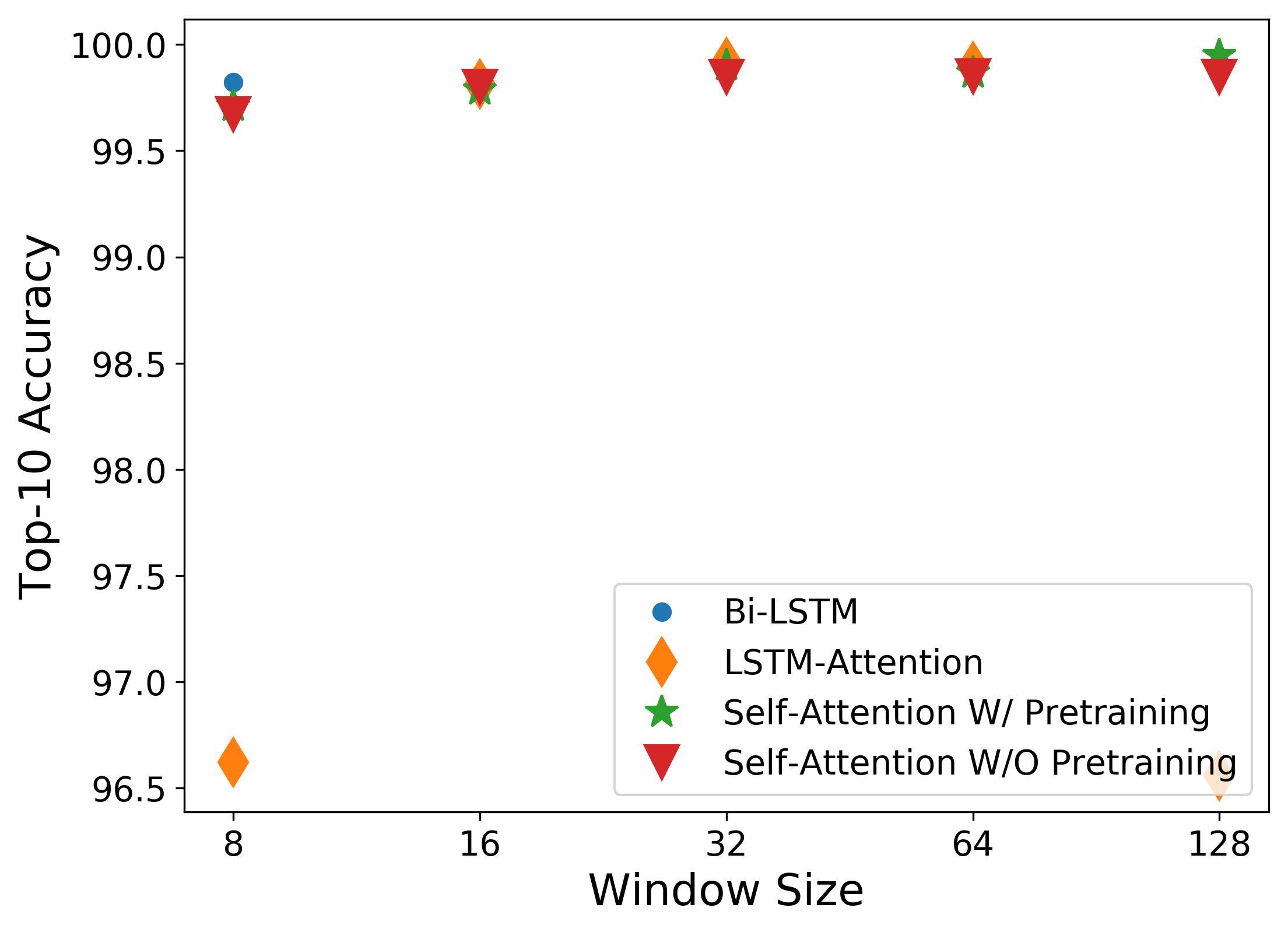}
\caption{DataRepo2}
\end{subfigure}
\caption{Model Comparison With Different Settings (Window Size and Pre-training).}
\label{fig:setting_online}
\vspace{-20pt}
\end{figure*}



\textbf{Impact of window size.}
To evaluate the impact of window sizes, we test five different window sizes of contextual events: 8, 16, 32, 64, 128. 
As illustrated in Figure~\ref{fig:setting_online}, self-attention based model performs better than Bi-LSTM and LSTM-attention based models for longer sequences (\ie, 128). The performance of Bi-LSTM and LSTM-attention based models degrade when the window size becomes 128. Bi-LSTM and LSTM-attention based models have decreasing weights on long-distance events as opposed to short-distance events, which may lead to vanishing gradients for longer-distance context. 

\textbf{Effectiveness of pre-training.}
Pre-training helps to learn not only the last-event prediction but also the semantic information and relationship of all events among the sequence.
We compare the performance of \bp with and without pre-training. As shown in Figure~\ref{fig:setting_online}, the performance is improved with the proposed pre-training technique for all six applications in general. On average, self-attention with pre-training increases Top-10 prediction accuracy compared to self-attention without pre-training. 

Especially, web applications with a large number of unique events may raise more uncertainty and lower accuracy in prediction. 
Pre-training largely improves the accuracy of prediction for these applications. 
For instance, the Top-10 accuracy of the three applications with the largest number of unique events (\ie, DevOpsApp, DataAnalyzer1, and DataAnalyzer2) is greatly increased by applying pre-training models on the self-attention based models.


\label{sec:center}
\begin{table}[!tb]
\centering
\caption{Performance of Predicting Centered Events.}
\label{tab:centered}
\begin{tabular}{llrr}
\toprule
\textbf{Application}           & \textbf{Model}     & \textbf{Top-1 Accuracy (\%)} & \textbf{Top-10 Accuracy (\%)} \\ \hline
\multirow{3}{*}{Workqueue}     & Bi-LSTM            & 97.07         & 99.79 \\  
                               & LSTM-Attention     & 97.27         & 99.81 \\ 
                               & Self-Attention & \textbf{98.33}         & \textbf{99.91}\\ \hline
\multirow{3}{*}{DataRepo1}     & Bi-LSTM            & 89.35         & 98.44\\
                               & LSTM-Attention     & 89.78         & 98.06\\
                               & Self-Attention & \textbf{90.33}         & \textbf{99.16}\\ \hline
\multirow{3}{*}{DevOpsApp}     & Bi-LSTM            & 74.00         & 91.72 \\ 
                               & LSTM-Attention     & 71.85         & 89.41 \\ 
                               & Self-Attention & \textbf{81.25}         & \textbf{94.67}\\ \hline
\multirow{3}{*}{DataAnalyzer1} & Bi-LSTM            & 84.31         & 96.00 \\ 
                               & LSTM-Attention     & 81.70         & \textbf{96.78} \\ 
                               & Self-Attention & \textbf{85.01}         & 96.61\\ \hline
\multirow{3}{*}{DataAnalyzer2} & Bi-LSTM            & 86.87         & 97.93\\ 
                               & LSTM-Attention     & 86.83         & 97.71\\ 
                               & Self-Attention & \textbf{87.79}         & \textbf{98.57}\\ \hline
\multirow{3}{*}{DataRepo2}     & Bi-LSTM            & 96.87         & 99.87\\ 
                               & LSTM-Attention     & 97.23         & 99.90\\
                               & Self-Attention & \textbf{97.96}         & \textbf{99.96}\\ \bottomrule
\end{tabular}
\vspace{-20pt}
\end{table}
\textbf{Evaluation of predicting centered events.}
In the previous experiments, we predicted the last event in a sequence. For many web applications, requests are generated concurrently by a single action. 
The concurrent requests make it possible for us to leverage contextual events following the event of interest. We study the case where the event of interest to be predicted is centered by contextual events.  

Table~\ref{tab:centered} shows the performance of \bp predicting centered events. Comparing Table~\ref{tab:model_comparison} and~\ref{tab:centered}, we observe that the prediction performance of centered events is improved for all three models in general. For example, for Workqueue, the Top-1 accuracy achieved by self-attention based model increases from 75.21\% to 98.33\%. Self-attention based model achieves more improvement than Bi-LSTM and LSTM-attention based models when the event of interest is centered by contextual ones. 
The performance of prediction improves significantly when we predict the centered event instead of the last one. When predicting centered event, events located after the event of interest provide important information. In this way, the model incorporates context from both directions (i.e., left and right). 
On average of six applications, \bp reduces Top-1 error rate by 52.56\% and Top-5 error rate by 57.84\% for predicting centered events.



\section{Related Work}


\vspace{-0.3cm}
\subsection{Web Event Forecasting}
\vspace{-0.2cm}
Web event forecast has been investigated for many years. Su \etal extracted access path from server logs and used n-gram models to predict web events for web caching and prefetching~\cite{su2000whatnext}. Awad \etal analyzed various supervised machine learning approaches for forecasting web events, such as Support Vector Machine, Markov Model and its variant, All-Kth Markov Model~\cite{awad2008predicting,awad2012prediction}. 
Da \etal summarized several clustering and Markov-based approaches for predicting web page access~\cite{da2018survey}.
In this work, we target the enterprise web applications and demonstrates superior performance in forecasting web events compared to the existing approaches.
\vspace{-0.3cm}
\subsection{Web Anomaly Detection}
\vspace{-0.2cm}
Many statistical models have been used to detect anomaly for web applications~\cite{kruegel2003anomaly,et2004anomaly,robertson2006using}. Kruegel \etal \cite{kruegel2003anomaly,kruegel2005multi} leveraged statistical models for characterizing HTTP query attributes such as query attribute length, attribute character distribution, and etc. Statistical models output probability values of a query and its individual attributes. The probability values reflect the likelihood of the occurrence with respect to an established profile. 
Juan \etal conducted Kruskal-Wallis and Kolmogorov-Smirnov test on payload length and payload histogram and modeled payload of normal web requests using Markov Chain~\cite{et2004anomaly}. 
Sakib and Huang detected HTTP-based Botnet C\&C traffic based on features from web request URLs and DNS responses. Three anomaly detection methods were used in the detection system: Chebyshev's Inequality, One-class SVM, and Nearest Neighbor based Local Outlier Factor.   
Many supervised machine learning provides have been used to detect anomaly for web applications by providing a binary prediction of normal or abnormal web requests learning from the historical data. Pham \etal surveyed different machine learning algorithms such as random forest, logistic regression, decision tree, AdaBoost, and SGD that are used to build Web intrusion detection systems~\cite{pham2016anomaly}. Oprea \etal detected malware in enterprises based on malicious HTTP traffic~\cite{oprea2018made}. 
They leveraged 89 features extracted from enterprise networks and applied several supervised machine learning algorithms (\eg, logistic regression, decision trees, random forest, and SVM) to learn from these features. 
Clustering and dimension-reduction are common techniques used in unsupervised learning based solutions~\cite{sipola2011anomaly,juvonen2012adaptive,Juvonen12015anomaly}. These solutions first extracted features from HTTP GET parameters and URLs, and then used Random Projection (RP), Principal Component Analysis (PCA), and Diffusion Map (DM) to reduce the dimensionality of the data. Clustering algorithms (\eg, K-means) have been applied to identify abnormal behavior.
Zolotukhin \etal \cite{zolotukhin2014anomaly} used several unsupervised learning algorithms such as PCA, K-means, Density-Based Spatial Clustering (DBSCAN) to model URL and User-Agent in HTTP headers and detect anomalies in web requests.

Recently deep learning approaches, in particular RNNs, have been established as state-of-the-art approaches in anomaly detection tasks. 
Liang \etal considered URLs as natural language sequences and applied LSTM and GRU to classify URLs as normal or abnormal requests~\cite{liang2017anomaly}. Yu \etal proposed a neural network consisting of Bidirectional LSTMs and an attention model to extract critical components from URI path and body~\cite{yu2018deephttp}. Liu \etal proposed an attention-based deep neural network, which located the malicious regions from HTTP requests and posts, and classified the malicious HTTP requests~\cite{liu2019locate}. 
These approaches focus on analyzing the contents in a single web request. We focus on a sequence of web requests, which involves connections among requests and represents users' normal patterns and web application flow characteristics.
\vspace{-0.3cm}
\subsection{Deep Neural Networks for Log Data Analysis}
\vspace{-0.2cm}
Deep neural networks have been used to analyze log data.
Du \etal proposed to model a sequence of system logs using LSTM and identified abnormal logs from normal execution~\cite{deeplog17}. An abnormal event is flagged if such an event is not within top-K probabilities to appear next. 
Shen \etal leveraged RNNs to predict future events based on previous observations using security logs collected from an intrusion prevention system~\cite{shenccs18}. The work focuses on the prediction of the upcoming security event given a sequence of events. 
Recently, Recurrent Neural Networks (RNNs) and its variants, Long Short-Term Memory (LSTM) \cite{lstm} and gated recurrent neural networks \cite{gatedrnn}, have been established as compelling techniques in security analytics research. 
The RNN based methods analyze the behavior of security event logs or system logs in a session. However, applying these models to web anomaly detection is non-trivial. Logs generated by machines (\eg, heartbeat) are much easier to be detected and predicted compared to web events generated by humans due to human's unpredictable behaviors.
To analyze web events, we adapt a self-attention mechanism to learn from the contextual events. With the proposed event and sequence embedding techniques, the adapted self-attention mechanism captures the dependency of long-distance events from human behaviors. 
\vspace{-0.4cm}


\section{Conclusion}
\vspace{-0.3cm}
In this work, we proposed a self-supervised neural network based approach, \bp for web event forecasting and anomaly detection.
We evaluated \bp on web requests collected from real-world enterprise web applications. 
By connecting web event forecasting with anomaly detection, \bp outperformed baseline methods and improved the performance of web event forecasting for complicated web events, while detected anomalies by identifying the most unlikely events in the sequence.
We also demonstrated \bp's capability in distinguishing normal web events from different types of anomalous events and measuring their anomaly scores.
\vspace{-0.3cm}

\bibliographystyle{splncs04}
\bibliography{deep.bib,security.bib,web.bib}


\end{document}